\documentclass[11pt,usletter]{article}
\usepackage{jheppub}
\usepackage{verbatim}
\usepackage{graphicx}
\usepackage{hyperref}
\usepackage{color}
\usepackage{amsmath}
\usepackage{tikz}
\usepackage{xcolor}

\usepackage[OT2,T1]{fontenc}
\DeclareSymbolFont{cyrletters}{OT2}{wncyr}{m}{n}

\title{ Ising Field Theory in a magnetic field: \\
$\varphi^3$ coupling at $T > T_c$}
\author{Hao-Lan Xu,}
\author{and Alexander Zamolodchikov}

\affiliation{C.N. Yang Institute for Theoretical Physics, State University of New York, Stony Brook, NY 11794-3840, USA}

\emailAdd{hao-lan.xu@stonybrook.edu}
\emailAdd{alexander.zamolodchikov@stonybrook.edu}
\begin{document}

\begin{flushright}
YITP-SB-2023-05\\
\end{flushright}
\abstract{
We study the "three particle coupling" $\Gamma_{11}^{1}(\xi)$, in $2d$ Ising Field Theory in a magnetic
field, as the function of the scaling parameter $\xi:=h/(-m)^{15/8}$, where $m \sim T_c-T$ and $h \sim H$ are scaled
deviation from the critical temperature and scaled external field, respectively. The "$\varphi^3$ coupling" $\Gamma_{11}^1$ is
defined in terms of the residue of the $2 \to 2$ elastic scattering amplitude at its pole associated with the lightest particle itself. We limit attention to the High-Temperature domain, so that $m$ is negative. We suggest "standard analyticity":
$(\Gamma_{11}^1)^2$, as the function of $u:=\xi^2$, is analytic in the whole complex $u$-plane except for the branch
cut from $-\infty$ to $-u_0 \approx -0.03585$, the latter branching point $-u_0$ being associated with the Yang-Lee edge singularity. Under this assumption, the values of $\Gamma_{11}^1$ at any complex $u$ are expressed through the discontinuity across the branch cut. We suggest approximation for this discontinuity which accounts for singular expansion of $\Gamma_{11}^1$
near the Yang-Lee branching point, as well as its known asymptotic at $u\to +\infty$. The resulting dispersion relation agrees well with known exact data, and with numerics obtained via Truncated Free Fermion Space Approach. This work is part of extended project of studying the S-matrix of the Ising Field Theory in a magnetic field.
}

\maketitle
\flushbottom

\section{Introduction}

In this work we report the next step in the extended project of studying the analytic properties of the thermodynamic and correlation characteristics of the 2d Ising Field Theory (IFT) in a magnetic field, as the functions of complex scaling parameter. We understand the IFT as the Ising conformal field theory (CFT) perturbed by its two relevant operators, the energy density $\varepsilon(x)$ and the spin density $\sigma(x)$, as defined by the formal action
\begin{eqnarray}\label{IFT}
\mathcal{A}_\text{IFT} = \mathcal{A}_\text{CFT} + \frac{m}{2\pi}\,\int \varepsilon(x) d^2x + h\,\int \sigma(x) d^2 x \,.
\end{eqnarray}
In \cite{fonseca2003ising} the analyticity of free energy was analysed in both High- and Low-Temperature regimes. Analyticity of the correlation length $R_c :=M_1^{-1}$ ($M_1$ stands for the mass of the lightest particle of the theory) was addressed in more recent work \cite{Xu:2022mmw}, albeit only in the High-T regime (The Low-T regime turns out to be more difficult; we plan to return to this problem in the future.) It was argued in \cite{Xu:2022mmw} that the correlation length, as the function of complex scaling parameter, is analytic on the complex plane of this parameter, the only singularity being the so called Yang-Lee edge singularity, the branching point point resulting from condensation of the Yang-Lee zeros in the thermodynamic limit. We refer to this analytic property as the "standard analyticity".

In this work we extend the analysis of \cite{Xu:2022mmw} to the "$\varphi^3$ coupling - the quantity which controls the large-distance behaviour of the three point correlation function. As in \cite{Xu:2022mmw}, we limit attention to the High-T domain. We argue that the $\varphi^3$ coupling also enjoys the standard analyticity. Our approach here is very similar to to the analysis in \cite{Xu:2022mmw}, and we often borrow notations from that paper.

Clearly, the Yang-Lee edge singularity plays the central role in our study. We refer the reader to the works \cite{Johnson:2022cqv}, \cite{Lencses:2022ira}, \cite{Klebanov:2022syt}, \cite{Cardy:2023lha} for recent developments in this area.

\subsection*{$\boldsymbol{\varphi}^3$ coupling}

We denote $A_1$ the lightest stable particle of the theory, and $M_1$ its mass. Also, as in \cite{Xu:2022mmw}, we use the notation $S_{11}(\theta)$ the $S$-matrix element of the $A_1+A_1 \to A_1+A_1$ elastic scattering. It is a function of the rapidity difference $\theta$. In IFT  $S_{11}(\theta)$ always has a pole at $\theta=2i\pi/3$ corresponding to the particle $A_1$ itself in the direct channel. (There is of course the associated cross-channel pole at $\theta = i\pi/3$.) The $\varphi^3$ coupling $\Gamma_{11}^1$ is usually defined in terms of the residue
\begin{eqnarray}
(\Gamma^1_{11})^2 = -i\, \underset{\theta=\frac{2i\pi}{3}}{\text{Res}} S_{11}(\theta) \,; \label{Residue}
\end{eqnarray}
by this definition $(\Gamma^1_{11})^2$ is positive in a unitary theory. The vertex $\Gamma^1_{11}$ is dimensionless. It depends on the scaling parameter, the dimensionless ratio of the coupling constants $m$ and $h$ in \eqref{IFT}. As in Refs.\cite{fonseca2003ising,Xu:2022mmw}, we use here the ratio $\xi = h/|m|^{15/8}$ (in the High-T domain $m$ is negative). We also use the related parameter $\eta = m/|h|^{8/15} = -\xi^{-8/15}$ when convenient. Because in the High-T regime IFT is unsensitive to the change of the sign of $\xi$ (the change $h\to-h$ in \eqref{IFT} can be compensated by the field transformation $\sigma\to-\sigma$), $(\Gamma^1_{11})^2$ in \eqref{Residue} is an even function of $\xi$\,\footnote{To the contrary, the vertex $\Gamma^1_{11}$ itself is an odd function, $\Gamma^1_{11}(-\xi)=-\Gamma^1_{11}(\xi)$. The sign of $\Gamma^1_{11}$ is fixed by the condition that $\Gamma^1_{11}$ is positive at positive $\xi$.}, and it is convenient to regard it as function of the variable $u:=\xi^2 = h^2/(-m)^{15/8}$. In what follows we use abbreviated notation $\Gamma:=\Gamma_{11}^1$, and study
the closely related function
\begin{eqnarray}\label{kappadef}
\kappa(u):=-\frac{\sqrt{3}}{2}\Gamma^2(\xi) = i\,\frac{\sqrt{3}}{2}\,\underset{\theta=\frac{2i\pi}{3}}{\text{Res}} S(\theta)\,.
\end{eqnarray}
Here and below $S(\theta)$ is the short-hand for $S_{11}(\theta)$.

The function $\kappa(u)$ can be analytically continued to complex values of $u$, and understanding the analytic properties of this continuation is the primary goal of the present work. As in \cite{Xu:2022mmw}, our analysis here will be based on the combination of exact results at the integrable points, as well as on the numerical data obtained via Truncated Free Fermion Space Approach (TFFSA) \cite{fonseca2003ising}.

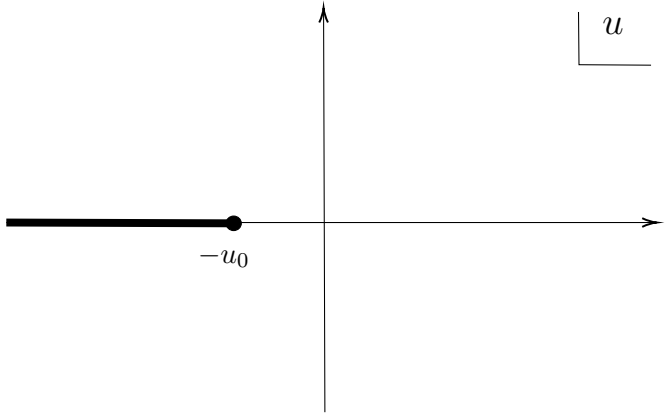
\begin{figure}[ht]
\centering
\begin{tikzpicture}[x=0.5pt,y=0.5pt,yscale=-1,xscale=1]
%uncomment if require: \path (0,896); %set diagram left start at 0, and has height of 896

%Straight Lines [id:da3399239833374874]
\draw    (79.8,671.8) -- (569.8,671.8) ;
\draw [shift={(571.8,671.8)}, rotate = 180] [color={rgb, 255:red, 0; green, 0; blue, 0 }  ][line width=0.75]    (10.93,-3.29) .. controls (6.95,-1.4) and (3.31,-0.3) .. (0,0) .. controls (3.31,0.3) and (6.95,1.4) .. (10.93,3.29)   ;
%Straight Lines [id:da6214634050687411]
\draw    (320.8,815.4) -- (319.81,511.4) ;
\draw [shift={(319.8,509.4)}, rotate = 449.81] [color={rgb, 255:red, 0; green, 0; blue, 0 }  ][line width=0.75]    (10.93,-3.29) .. controls (6.95,-1.4) and (3.31,-0.3) .. (0,0) .. controls (3.31,0.3) and (6.95,1.4) .. (10.93,3.29)   ;
%Straight Lines [id:da7599973469533079]
\draw [line width=3]    (79.8,671.8) -- (251.8,672.4) ;
%Flowchart: Summing Junction [id:dp8744190220896932]
\draw   [fill={rgb, 255:red, 0; green, 0; blue, 0 }  ,fill opacity=1 ] (246.02,672.4) .. controls (246.02,669.36) and (248.61,666.9) .. (251.8,666.9) .. controls (254.99,666.9) and (257.58,669.36) .. (257.58,672.4) .. controls (257.58,675.44) and (254.99,677.9) .. (251.8,677.9) .. controls (248.61,677.9) and (246.02,675.44) .. (246.02,672.4) -- cycle ; \draw   (247.71,668.51) -- (255.89,676.29) ; \draw   (255.89,668.51) -- (247.71,676.29) ;
%Shape: Right Angle [id:dp5958880712746788]
\draw   (567.56,552.67) -- (513.04,552.31) -- (512.67,512.39) ;

% Text Node
\draw (223,687.8) node [anchor=north west][inner sep=0.75pt]    {$-u _{0}$};
% Text Node
\draw (529,514.67) node [anchor=north west][inner sep=0.75pt]  [font=\Large]  {$u$};
\end{tikzpicture}
\caption{Conjectured "standard analyticity" of $\kappa(u)$ in the complex $u$-plane: $\kappa(u)$ is analytic everywhere
except for the branch cut $[-\infty: -u_0]$. The branching point $-u_0$ is the YL edge singularity. The branch cut represents the shown as the solid line represents the line of first order phase transitions. In particular, the $\varphi^3$ coupling is discontinuous across this line.}
\label{uplaneFig}
\end{figure}

It is natural to expect that analytic properties of $\kappa(u)$ in the complex
$u$-plane are similar to those of the vacuum energy density $G_\text{high}(u)$  \cite{fonseca2003ising}, and of the mass $M_1(u)$ of the lightest particle $A_1$ \cite{Xu:2022mmw}. Namely, one can argue that $\kappa(u)$ is analytic in the whole $u$-plane with the branch cut from $-\infty$ to $-u_0 = -\xi_0^2 \approx -0.03585\dots$, as shown in Fig.\ref{uplaneFig}. We refer to it as the "standard analyticity". The branching point $u=-u_0$ represents the Yang-Lee edge singularity resulting
from accumulation of Yang-Lee zeros in the thermodynamic limit\cite{yang1952statistical,lee1952statistical}. The argument
is as follows. Generally, the positions and the residues of the poles of $S(\theta)$ may have, besides the singularity at $-u_0$, additional algebraic singularities (square root branching points) signifying collisions
of different poles at special values of $u$, with two branches of the square root representing interchange of the colliding poles. However, the particular pole at $\theta=2\pi i/3$ which we are interested in here stays put at all values of $u$. There is an interesting movements of other poles and associated zeros under changes of $u$, as explained in Ref.\cite{zamolodchikov2013ising}. The choreography of those movements is such that the residue \eqref{Residue} never does evolve singularities. Every time another pole hits the pole at $2\pi i/3$, there is a zero hitting it simultaneously, so that the residue remains regular.

Below we verify this analyticity conjecture using a combination of exact
and numerical data. We build an approximation for the discontinuity across the above branch cut. Then the analyticity assumption leads to the dispersion relation, and one can check it against special points where $\kappa(u)$ is known exactly, as well as the numerical estimates obtained using TFFSA. We find reasonable agreement (see Sec.4 and Sec.5).
Alternative numerical data for $\kappa(u)$ at real positive $u$ was previously obtained in \cite{Gabai:2019ryw} by rather different approach. Our results from the dispersion relation are in good agreement with that numerics as well. All these results support the standard analyticity conjecture formulated above.

\section{Special points}

The theory \eqref{IFT} is exactly integrable at two special points. At $u=0$ it reduces to a theory of free fermi particles with
$S(\theta)=-1$, so that the residue \eqref{Residue} vanishes. Away from this point $\kappa(u)$ admits expansion in powers of $u$,
\begin{eqnarray}\label{kappaexpu}
\kappa(u) = \sum_{n=1}^\infty\, \kappa^{(n)}\,u^n\,,
\end{eqnarray}
with finite radius of convergence. The leading term in this expansion was obtained in \cite{Zamolodchikov:2011wd} via form factor perturbation theory,
\begin{eqnarray}\label{kappauslope}
\kappa^{(1)} = -18\sqrt{3} \, {\bar s}^2 = -57.48165545 \dots \,,
\end{eqnarray}
where ${\bar s} = 2^{1/12} e^{-1/8} \,A_G^{3/2} = 1.35783834...$, and $A_G$ is Glaisher's constant\cite{mccoy2014two}.

Another important point is $u=\infty$ (i.e. $m=0$), where the \eqref{IFT} becomes an integrable theory with eight stable
particles and rich factorizable S-matrices \cite{zamolodchikov1989integrals}\footnote{The masses of the eight particles $A_p$, $p=1,2,...,8$ at this integrable point are proportional to the components of Frobenius vector of the Cartan matrix of the
Lie algebra $E_8$, and the structure of the S-matrices reflects in many ways the
properties of the $E_8$ root system \cite{zamolodchikov1989integrals}. For this reason we refer to this
integrable theory as the "$E_8$ theory".}. The value of $\kappa(u)$ at $u=\infty$ can be straightforwardly extracted from
the S-matrix $S_{11}(\theta)$ (Eq.\eqref{S11-E8} in Appendix A),
\begin{eqnarray}\label{binfinity}
\kappa(\infty) = -3\,\frac{\tan(7\pi/15)\tan(11\pi/30)}{\tan(2\pi/15)\tan(3\pi/10)} = -104.6154448\dots \,.
\end{eqnarray}
In the vicinity of this point $\kappa(u)$ expands in a convergent series in integer powers of $\eta= -u^{-4/15}$,
\begin{eqnarray}
\kappa(u) = \kappa_0 + \kappa_1\,\eta + \kappa_2\,\eta^2 + \kappa_3\,\eta^3 + \dots \label{kappaexpansioneta}
\end{eqnarray}
where $\kappa_0:=\kappa(\infty)$, see Eq.\eqref{binfinity}. In principle, the higher coefficients can be obtained via form factor
perturbation theory around the integrable point $u=\infty$.  In practice, this can be done for $\kappa_1$ (see Appendix A),
\begin{eqnarray}
\kappa_1 = - 202.04... \label{b1}
\end{eqnarray}
while exact calculations of the higher coefficients are rather difficult. We will estimate some coefficients using numerical data in Sec.4 and Sec.5.

Interesting point where IFT \eqref{IFT} is "infrared integrable" is the Yang-Lee edge singularity which is located at pure
imaginary magnetic field $h$, i.e. at real negative $u=-u_0$. The position $u_0$ was estimated in \cite{fonseca2003ising,BazhanovYL,Xu:2022mmw},
$u_0 \approx 0.035846(4)$. This point is critical in usual
sense - the analytic continuation of the mass $M_1(u)$ vanishes at $u=-u_0$, i.e. the correlation length diverges
\cite{fisher1978yang}. This means that the deep infrared asymptotic of the theory \eqref{IFT} at $u=-u_0$ is described by
CFT, which was identified in \cite{cardy1985conformal} as the non-unitary minimal CFT $\mathcal{M}_{2/5}$ (in what follows we refer to it as
YLCFT). Away from the YL point the theory is gapped, with the mass $M_1(u) \sim (u+u_0)^{5/12}$, while its deep IR asymptotic
is described by the so called Yang-Lee QFT (henceforth referred to as YLQFT),
\begin{eqnarray}\label{YLQFT}
\mathcal{A}_\text{YLQFT} = \mathcal{A}_\text{YLCFT} + i\lambda(u)\,\int \varphi(x) d^2 x\,,
\end{eqnarray}
where $\varphi(x)$ is the only nontrivial scalar primary with the conformal dimensions $(-\frac{1}{5},-\frac{1}{5})$ of the YLCFT. The coupling constant
$\lambda=\lambda(u)$ turns to zero at the YL point, $\lambda(-u_0)=0$, in a regular way $\lambda(u) \sim (u+u_0)$. The YLQFT \eqref{YLQFT} itself is known to be integrable, and its factorizable S-matrix is determined by the $2\to 2$ elastic amplitude \cite{Cardy:1989fw}
\begin{eqnarray}\label{SmatrixYL}
S_{11}^{(\text{YLQFT})}(\theta) = \frac{\sinh\theta +i\sin(2\pi/3)}{\sinh\theta - i\sin(2\pi/3)}\,.
\end{eqnarray}
However, the full theory \eqref{IFT} is not integrable even in small vicinity of the YL point, not even at the YL point
$u=-u_0$ itself. While at $u$ close to $-u_0$ \eqref{YLQFT} describes deep IR asymptotic of the theory \eqref{IFT}, in the full
theory the YLQFT is dressed with an infinite tower of irrelevant operators,
\begin{eqnarray}\label{aeff0}
\mathcal{A}_\text{IFT}\ \to\ \mathcal{A}_\text{YLQFT} + \sum_i\,a^i \int O_i(x) d^2 x\,,
\end{eqnarray}
which become visible at intermediate scales $R \lesssim M_1^{-1}$. Here $O_i$ are scalar operators of YLCFT with
$\Delta_i = \bar \Delta_i >1$; contributions of these operators generally break integrability \cite{Xu:2022mmw}
(We will say more about these operators in Sec.3 below).
While $M_1$ measured in the units of $|m|$ goes to zero when $u\to-u_0$, the couplings $a^i = a^i(u)$ remain finite in this limit.
Equivalently, if measured in the units of $M_1$, these couplings tend to zero when $u\to u_0$. Therefore, the value of $\kappa$
at $u=-u_0$ can be read out of the S-matrix \eqref{SmatrixYL},
\begin{eqnarray}\label{kappaYL}
\kappa(-u_0) = i\frac{\sqrt{3}}{2}\,\underset{\theta=\frac{2i\pi}{3}}{\text{Res}}\,S_{11}^\text{YLQFT}(\theta) = 3.
\end{eqnarray}
The negative value of $\Gamma^2$ at this point reflects non-unitary nature of YLQFT.

\section{Singular expansion and TTbar deformation}

In the vicinity of the YL point $\kappa(u)$ admits singular expansion in fractional
powers of $u+u_0$, generated by the irrelevant operators $O_i$ in the effective action \eqref{aeff0}. Schematically, the structure of the expansion is
\begin{eqnarray}\label{kappasing0}
\kappa(u) = \sum_{n=1}^\infty\ \sum_{i_1, ..., i_n}\, H_{i_1, i_2, ...,i_n}\,a^{i_1}(u)...a^{i_n}(u)\,
\left[ M_1(u)\right]^{\delta_{i_1} + ... + \delta_{i_n}}
\end{eqnarray}
where $-\delta_i = 2( 1- \Delta_i) < 0$ are the mass dimensions of the couplings $a^i$, and $H_{i_1,...,i_n}$ are dimensionless coefficients. While the couplings $a^i(u)$ are regular at $u=-u_0$,
\begin{eqnarray}
a^i (u) = a_0^{i} + a_1^{i}\,(u+u_0) + a_2^i\,(u+u_0)^2 + ...
\end{eqnarray}
the mass $M_1(u)$ turns to zero at $u=-u_0$ in a singular manner, $M_1(u)\sim [\lambda(u)]^{5/12} \sim (u+u_0)^{5/12}$, and further admits expansion in fractional powers of $(u+u_0)$ around the YL point, see \cite{Xu:2022mmw}, and Eq.\eqref{Mexp} below. The terms in \eqref{kappasing0} can be interpreted as the contributions from the perturbative expansion in the couplings $a^i(u)$. Of course, the perturbation theory in the irrelevant operators is non-renormalizable, and by itself it does not unambiguously define the coefficients in \eqref{kappasing0} beyond the linear order in the couplings $a^i$. (Important exception is the "least irrelevant" operator $T{\bar T}$ whose contributions are unambiguous to all orders in the associated coupling $\alpha$; we discuss its role below.) Nonetheless, \eqref{kappasing0} shows the nature of the singular expansion of $\kappa(u)$ around the YL singularity.

The list of the scalar irrelevant operators entering the expansion \eqref{aeff0} can be extracted from the known operator content of YLCFT, aka the minimal CFT
$\mathcal{M}_{2/5}$. We arrange them in order of growing dimensions $\Delta_i$, and exclude total derivatives as they do not contribute to the action \eqref{aeff0}. Then the lowest of the irrelevant operators is
\begin{eqnarray}\label{ttbardef}
(T{\bar T}):= L_{-2}{\bar L}_{-2}I\,, \qquad\ \ \ (\Delta,{\bar\Delta}) = (2,2)\,.
\end{eqnarray}
It is well known to generate the so-called TTbar deformation of the YLQFT, which makes it possible to treat its contributions to all orders (see \cite{smirnov2017space,Cavaglia:2016oda}), as we will discuss shortly. Let us also display the next three operators,
\begin{eqnarray}\label{xidef}
&&\Xi:=L_{-4}{\bar L}_{-4}\varphi\,, \qquad \qquad (\Delta,{\bar\Delta})=(3.8, 3.8)\,,\\
&&\Xi_6 = L_{-6}{\bar L}_{-6}\varphi\,, \qquad\qquad (\Delta, {\bar\Delta})=(5.8,5.8)\,,\\
&&(T{\bar T})^3 := L_{-2}^3{\bar L}_{-2}^3 I\,, \ \qquad (\Delta, {\bar\Delta}) = (6,6)\,.\label{ttbar6def}
\end{eqnarray}
Thus, the least irrelevant terms in \eqref{aeff0} are
\begin{eqnarray}\nonumber
&&\eqref{aeff0} = \mathcal{A}_\text{YL QFT} + \frac{\alpha(u)}{\pi^2}
\int (T{\bar T})(x) d^2 x + \frac{\beta(u)}{2\pi}\,\int\Xi(x) d^2 x +\\
&&\qquad\qquad \qquad\frac{\gamma(u)}{2\pi}\int \Xi_6(x) d^2 x + \frac{\alpha_5(u)}{\pi^2}\int (T{\bar T})^3 (x) d^2 x\ +\ ...\label{aeff1}
\end{eqnarray}
where we used individual notations for the couplings. Like all the couplings
in \eqref{aeff0}, the couplings $\alpha, \beta, \gamma$ in \eqref{aeff1} are functions of $u$, regular at the YL point; they
admit convergent expansions around the YL point,
\begin{eqnarray}\nonumber
&&\alpha(u) = \alpha_0 + \alpha_1\,(u+u_0) + \alpha_2\,(u+u_0)^2 + ...\,, \qquad \qquad -\delta_\alpha = -2\,,\\
&&\beta(u) = \beta_0 + \beta_1\,(u+u_0) + \beta_2\,(u+u_0)^2 + ...\,,\ \qquad \qquad -\delta_\beta = -5.6\,,\label{betaexp}\\
&&\gamma(u) = \gamma_0 + \gamma_1\,(u+u_0) + \gamma_2\,(u+u_0)^2 + ...\,,\qquad\qquad\ \, -\delta_\gamma = -9.6\,,\nonumber\\
&&\alpha_5(u) = \alpha_{5,0} + \alpha_{5,1}\,(u+u_0) + \alpha_{5,2}\,(u+u_0)^2 + ...\,, \qquad -\delta_{\alpha_5} = -10\,,\nonumber
\end{eqnarray}
and similarly for further couplings not displayed in \eqref{aeff1}. In \eqref{betaexp} we also show the mass dimensions of the couplings. The leading coefficients for the first two coupling parameters were determined in \cite{Xu:2022mmw} (see also \cite{fonseca2003ising}),
\begin{eqnarray}\label{alphabeta0}
\alpha_0\,|m|^2 = -1.32(5)\,, \qquad \frac{\beta_0}{2\pi}\,|m|^\frac{28}{5} =+0.72(6)\,.
\end{eqnarray}
In Sec.5 below we use TFFSA to obtain numerical estimate (not very precise) $\alpha_1|m|^2 = 18.0(3.5)$ of the first sub-leading coefficient in \eqref{betaexp}. Estimates of higher coefficients in \eqref{betaexp} are not yet available.

With the account of the three lowest irrelevant operators displayed in \eqref{aeff1} the expansion \eqref{kappasing0} takes the
form
\begin{eqnarray}\nonumber
&&\kappa(u) = 3 + H_1\,\alpha(u) M_1^2(u) + H_2\,\alpha^2(u) M_1^4(u) + \\
&&H_1^\Xi\,\beta(u) M_1^\frac{28}{5}(u) +
H_3\,\alpha^3(u) M_1^6(u) + H_4\,\alpha^4 (u) M_1^8(u) + \label{kappaexp1}\\
&&\left({ H}_2^\Xi\,\alpha(u)\beta(u) + H_1^{\Xi_6}\,\gamma(u)\right)\,M_1^\frac{48}{5}(u) + \left(H_5\,\alpha^5(u) + {\tilde H}_1\,\alpha_5(u)\right) M_1^{10}(u) + ...\nonumber
\end{eqnarray}
where the leading term comes from the exact number \eqref{kappaYL}.
The omitted terms have the powers of $M_1(u)$ 11.6 and higher. Note that apart from the terms linear in the couplings, \eqref{kappaexp1} involves higher powers of $\alpha$, as well as the term $\alpha\beta$. These are
associated with the higher order perturbation theory in $\alpha$. Generally, such terms can not be determined within the perturbation theory itself, as they depend on short-distance behavior of the full IFT. However, due to special properties of the operator $(T{\bar T})$ many
such terms are uniquely defined. As was already mentioned, this operator generates the TTbar deformation, which allows to
collect explicitly the higher order terms in $\alpha$. We will not explain it here, referring the reader to \cite{smirnov2017space,Cavaglia:2016oda,Dubovsky:2017cnj}.
In particular, the TTbar deformation leads to simple "dressing" of the two-particle elastic S-matrix \cite{smirnov2017space,Dubovsky:2017cnj},
\begin{eqnarray}\label{sttbar}
S^{(\alpha)}(\theta) = \exp\left\{-i\alpha M_1^2\,\sinh\theta\right\}\,S^{(0)}(\theta)\,,
\end{eqnarray}
where $S^{(0)}(\theta)$ is the S-matrix of the undeformed theory\footnote{For "TTbar dressing" of the full S-matrix see \cite{Dubovsky:2017cnj}.}. If $S^{(0)}(\theta)$ has a pole at $\theta=2\pi i/3$, so does the deformed $S^{(\alpha)}(\theta)$, with the residue determined by \eqref{sttbar}. Therefore the "TTBar deformed" $\kappa^{(\alpha)}$
relates to $\kappa^{(0)}$ at $\alpha=0$ as
\begin{eqnarray}\label{kappattbar}
\kappa^{(\alpha)} = \exp\left\{\frac{\sqrt{3}}{2}\,\alpha\,M_1^2\right\}\,\kappa^{(0)}\,.
\end{eqnarray}

In our context only five leading terms of the expansion of the exponential in \eqref{sttbar} and \eqref{kappattbar} are useful.
This is because the higher order terms interfere with the contributions of higher level descendants of the identity operator in the
effective action \eqref{aeff1}. Thus, the term $\sim\alpha^5$ mixes with the leading contribution from $(T{\bar T})^3$
($\alpha^5$ has the same mass dimension as $\alpha_5$)\footnote{From the point of view of straightforward perturbation theory, the TTbar deformation itself generates an infinite tower of higher descendants of $I$, which mix with the higher terms in \eqref{aeff1}.
Alternatively, one can think of the term $\alpha(T{\bar T})$ as the "germ" of the full TTbar deformation; with this convention
\eqref{sttbar} is exact to all orders in $\alpha$. Likewise, the term $\alpha_5 (T{\bar T})^3$, if understood as the germ
of generalized TTbar deformation \cite{smirnov2017space}, leads to exact dressing by the factor
$$
\exp\left\{-i\alpha M_1^2\sinh(\theta)-i\alpha_5 M_1^{10}\sinh(5\theta) + ...\right\}
$$
}.
For the purpose of this work, the expansion
\begin{eqnarray}\label{kappaexp3}
\kappa^{(\alpha)} = \left(1 + \frac{\sqrt{3}}{2}\alpha M_1^2 + \frac{3}{8}\alpha^2 M_1^4 + \frac{\sqrt{3}}{16}\alpha^3 M_1^6
+ \frac{3}{128}\alpha^4 M_1^8\right) \,\kappa^{(0)} + O(\alpha^5)
\end{eqnarray}
is more than sufficient. Here $\kappa^{(0)}$ is given by the expansion \eqref{kappaexp1} with $\alpha(u)$ set to zero,
\begin{eqnarray}\label{kappa0exp}
\kappa^{(0)}(u) = 3 + H_1^\Xi\,\beta(u)\,M_1^\frac{28}{5}(u) + H_1^{\Xi_6}\,\gamma(u)\,M_1^\frac{48}{5}(u) + O(M_1^{10})\,.
\end{eqnarray}

The second and the third terms in the r.h.s. of \eqref{kappa0exp} are linear in $\beta$ and $\gamma$; these terms express the leading order contributions from the operators $\Xi$ and $\Xi_6$ in \eqref{aeff1}. As usual, unlike the higher orders, these coefficients are determined unambiguously using the perturbation theory. The calculation involves the four particle form factors of the operators
$\Xi$ and $\Xi_6$, and is rather bulky. We will present detailed calculations elsewhere. Here we just quote the result \cite{XuEtAl2022},
\begin{eqnarray}
H^\Xi = - \frac{164480625}{100352} \frac{5^{\frac{1}{4}} \, \pi^{\frac{21}{5}} \, \Gamma(\frac{3}{5}) \, \Gamma(\frac{2}{3})^{\frac{38}{5}} \, \Gamma(\frac{4}{5}) \, \Gamma(\frac{8}{3})^2 }{2^{\frac{1}{5}} \, \Gamma(\frac{1}{6})^{10} \, \Gamma(\frac{5}{6})^{\frac{12}{5}} \, \Gamma(\frac{25}{6})^2 } = -0.004876389\dots \,.
\end{eqnarray}
One can note that this coefficient is exceedingly small numerically, and it does not have any noticeable effect on the analysis
in Sec.4 and Sec.5 below. We also have reasons to believe that $H^{\Xi_6}$ is yet smaller. Because of this, in what follows we
ignore the $\beta$- and $\gamma$-contributions in \eqref{kappa0exp}.

Important ingredient in \eqref{kappasing0} is the mass $M_1(u)$ which enjoys
the singular expansion around the YL point \cite{Xu:2022mmw},
\begin{eqnarray}\label{Mexp}
\frac{M_1(u)}{|m|} = (u+u_0)^\frac{5}{12}\left(b(u) + (u+u_0)^\frac{5}{6}\,c(u) + (u+u_0)^\frac{5}{3}\,e(u)  +
(u+u_0)^\frac{7}{3} \, d(u) + \dots \right)\,, \quad
\end{eqnarray}
where $b(u), c(u), d(u), e(u) ...$ are regular at $u=-u_0$,
\begin{eqnarray}\label{bexp}
&&b(u) = b_0 + b_1\,(u+u_0) + b_2\,(u+u_2)^2 + ...\\
&&c(u) = c_0 + c_1\,(u+u_0) + ...\label{cexp}
\end{eqnarray}
and similarly for $e(u)$, $d(u)$ and the higher coefficient functions. Some lower
coefficients in \eqref{bexp},\eqref{cexp} were determined in \cite{Xu:2022mmw},
\begin{eqnarray}\label{bbcu}
b_0 = 4.228(5), \qquad b_1 = 21.9(9)\,, \qquad c_0 = -14.4(6)\,.
\end{eqnarray}

For the Ising effective action near the YL point, Eq.\eqref{aeff1}, the expansion \eqref{kappasing0} combined with \eqref{betaexp}, generates expansion of $\kappa(u)$ in fractional powers of $u+u_0$\
\begin{eqnarray}\nonumber
&&\kappa(u) = 3 + k_1(u+u_0)^\frac{5}{6} + k_2(u+u_0)^\frac{5}{3}+ \qquad\qquad\qquad\qquad\qquad\qquad\\
&&\qquad\qquad\qquad\qquad k_3(u+u_0)^\frac{11}{6} + k_4(u+u_0)^\frac{5}{2} + k_5(u+u_0)^\frac{8}{3} +...\,,
\end{eqnarray}\label{kappauexp}
where the coefficients are expressed in terms of the coefficients of the expansion of $\alpha(u)$ in \eqref{betaexp}, as well as the coefficients in \eqref{Mexp}, e.g.
\begin{eqnarray}
&&k_1 = \frac{3\sqrt{3}}{2}\,b_0^2\,\alpha_0 \approx -61.3 \,,\\
&&k_2 = \frac{9}{8}\,b_0^2\,\alpha_0^2 \approx 35.0 \,,\\
&&k_3 =  \frac{3\sqrt{3}}{2}\,(2 b_0 b_1\,\alpha_0 +\alpha_1) \approx -681 \,,
\end{eqnarray}
where the estimate of $\alpha_0$, Eq.\eqref{alphabeta0}, is used to produce numerics for $k_1$ and $k_2$. Higher $k_n$ involve higher coefficients in \eqref{alphabeta0} and \eqref{Mexp}. We give numerical estimate
of more $k_n$ in Sec.4 and Sec.5.

\section{Analyticity and Dispersion relation}

According to our analyticity conjecture, the function $\kappa(u)$ admits analytic continuation to the complex $u$-plane, with
no singularities except for the branch cut along the real axis from $-\infty$ to $-u_0$, as shown in Fig.\ref{uplaneFig}. Under this assumption the values of
$\kappa(u)$ everywhere in the $u$-plane are expressed in terms of the discontinuity across the branch cut, $\text{Disc}\,\kappa(u): = \kappa(u+i0)-\kappa(u-i0) = 2i\,\Im m \,\kappa(u+i0)$
\begin{eqnarray}\label{Residue_dispersion_u}
\kappa(u) = -u\,\int_{u_0}^\infty\,\frac{dx}{\pi}\,\frac{\Im m \,\kappa(-x)}{x (x+u)}
\end{eqnarray}
where we have taken into account that $\kappa(0)=0$. The integral in \eqref{Residue_dispersion_u} converges because $\Im m\,\kappa(-x) \simeq x^{15/16}$ as $x\to +\infty$, see \eqref{kappaexpansioneta}. Here we are going to build certain approximation for the discontinuity
$\text{Disc}\,\kappa(u)$ and then check the numerics for $\kappa(u)$ obtained via \eqref{Residue_dispersion_u} against exact
numbers in Sec.2 and Sec.3. In addition, in the next section we use TFFSA to obtain numerical values of $\kappa(u)$ at generic real $u>-u_0$, and test the dispersion relation against that numerics as well.

We found it convenient to to use the scaling variable
\begin{eqnarray}
\eta = - u^{-4/15}
\end{eqnarray}
instead of $u$, and define
\begin{eqnarray}\label{dispu}
{\hat \kappa}(\eta) = \kappa \left((-\eta)^{-15/4}\right)\,.
\end{eqnarray}
The variable transformation maps the $u$-plane in Fig.\ref{uplaneFig} to the wedge
\begin{eqnarray}\label{etawedge}
-4\pi/15 < \text{arg}(-\eta)<4\pi/15
\end{eqnarray}
in the $\eta$-plane, with the positive real $u$ mapped to the negative part of the real axis of $\eta$, and negative real $u$
mapped to the rays $\eta=y\,e^{\pm 4\pi i/15}$ with negative real $y$. In these variable, the dispersion relation \eqref{Residue_dispersion_u} takes the form
\begin{eqnarray}\label{dispeta}
{\hat \kappa}(\eta) = -\frac{15}{4\pi}\,\int_{0}^{Y_0}\,\frac{y^\frac{11}{4}\,\Delta(-y)\,dy}{\big(y^{\frac{15}{4}}+(-\eta)^{\frac{15}{4}}\big)}
\end{eqnarray}
where it is assumed that $\eta$ lays in the wedge \eqref{etawedge}, and
\begin{eqnarray}
Y_0 = u_0^{-4/15} = 2.4293 \dots \,.
\end{eqnarray}
The function
\begin{eqnarray}\label{Deltadef}
\Delta(y) = \frac{1}{2i}\,\left[{\hat \kappa}\big(y\,e^{\frac{4\pi i}{15} + i0}\big)-{\hat \kappa}\big(y\,e^{\frac{4\pi i}{15} - i0}\big)\right] =
\Im m\, {\hat \kappa}\big(y\,e^{\frac{4\pi i}{15} + i0}\big)
\end{eqnarray}
represents the discontinuity across the branch cut in Fig.\ref{uplaneFig}. To make \eqref{dispeta} useful we need to build some
approximation for $\Delta(y)$.

\subsection*{Approximating $\boldsymbol{\Delta(y)}$}

As was mentioned in Sec.2, the function ${\hat\kappa}(\eta)$ admits convergent expansion \eqref{kappaexpansioneta} in integer powers
of $\eta$. This generates expansion of the function \eqref{Deltadef},
\begin{eqnarray}
\Delta(y) = \sum_{n=0}^\infty\,\kappa_n\,\sin\left(\frac{4\pi n}{15}\right)\,y^n = \kappa_1\,\sin\left(\frac{4\pi}{15}\right)\,y +
\kappa_2\,\sin\left(\frac{8\pi}{15}\right)\,y^2 + ...
\end{eqnarray}
The first two coefficients are
\begin{eqnarray}
\Delta(0)=0\,, \qquad \Delta'(0) = \kappa_1\,\sin\left(\frac{4\pi}{15}\right) = -150.14 \dots \,, \label{IntersectionSlope}
\end{eqnarray}
where we have used \eqref{b1}.

On the other hand, ${\hat\kappa}(\eta)$ enjoys the singular expansion near the YL point $y+Y_0=0$,
\begin{eqnarray}
{\hat\kappa}(y) = 3+\epsilon^\frac{5}{6}\,K_1(\epsilon)+\epsilon^\frac{5}{3}\,K_2(\epsilon) + \epsilon^{\frac{5}{2}} \, K_3(\epsilon) + \epsilon^{\frac{19}{6}} \, K_4(\epsilon)  + \epsilon^{\frac{10}{3}} \, K_5(\epsilon) + \epsilon^{\frac{25}{6}}\, K_6(\epsilon) + \cdots \,, \quad \label{kappasingYL}
\end{eqnarray}
Here and below we use the notation $\epsilon:=-(y+Y_0)$, so that
\begin{eqnarray}
u+u_0 = -(-y)^{-15/4}+Y_0^{-15/4} = \frac{15}{4}\,Y_0^{-\frac{19}{4}}\,\epsilon - \frac{285}{32} Y_0^{-\frac{23}{4}}\,\epsilon^2 + ...
\end{eqnarray}
The coefficient functions $K_n(\epsilon)$ in \eqref{kappasingYL} are regular around the YL point
\begin{eqnarray}
K_n(\epsilon) = K^{(0)}_n +  K^{(1)}_n\,\epsilon + K^{(2)}_n\,\epsilon^2  + \cdots \,.
\end{eqnarray}
Like in \eqref{kappauexp}, the coefficients $K_n^{(l)}$ are related in a straightforward way to the coefficients of the
expansion of the $(T{\bar T})$ coupling
\begin{eqnarray}\label{alphaepsilon}
\alpha = \alpha_0 + {\hat \alpha_1}\,\epsilon + {\hat\alpha}_2\,\epsilon^2 +...
\end{eqnarray}
and the mass,
\begin{eqnarray}\label{Mepsilon}
M_1/ |h|^{\frac{8}{15}} = \epsilon^\frac{5}{12}\,({\hat b}_0 + {\hat b}_1\,\epsilon + {\hat b}_2\,\epsilon^2) + \epsilon^\frac{5}{4}\,({\hat c}_0 +
{\hat c}_1\,\epsilon) + \epsilon^\frac{25}{12} {\hat e}_0  + \epsilon^\frac{11}{4} \,{\hat d}_0 + O(\epsilon^\frac{19}{6})\,,
\end{eqnarray}
which are obtained by re-expanding  \eqref{betaexp} and \eqref{Mexp} in terms of $\epsilon$. The coefficient $\alpha_0$ is the same
as in \eqref{alphabeta0} while
\begin{eqnarray}
{\hat\alpha}_1\,|m|^2 = - 0.85 \pm 0.1 \label{alpha1y}
\end{eqnarray}
will be determined in Sec.5. Moreover, analysis in Sec.5 suggests that ${\hat\alpha}_2$ is small, and within the (low) accuracy of
our numerics we will neglect its contributions. The coefficients ${\hat b}_0$, ${\hat b}_1$, and ${\hat c_0}$ in \eqref{Mepsilon} are expressed through the numbers \eqref{bbcu}, see corresponding entries in Tab.\ref{7ParametersSingularCoefficients}. Moreover, it was found in \cite{Xu:2022mmw} that the mass $M_1$ is approximated well at $y+Y_0 \lesssim 1$ by all six terms explicitly displayed in \eqref{Mepsilon}, with the numerical values of the coefficients in Tab.\ref{7ParametersSingularCoefficients}.

\begin{table}[htp]
\begin{center}
\begin{tabular}{|c|c|c|c|c|c|c|}
\hline
${\hat b}_0$ & ${\hat b}_1$ & ${\hat b}_2$ & ${\hat c}_0$ & ${\hat c}_1$ & ${\hat d}_0$ & ${\hat  e}_0$ \\
\hline
$3.0754$ & $0.8932$ & $-0.8618$ & $-0.9412$ & $0.7835$ & $0.3134$ & $  0.2880$\\
\hline
\end{tabular}
\end{center}
\caption{Numerical values of the expansion coefficients in Eq.\eqref{Mepsilon}.}
\label{7ParametersSingularCoefficients}
\end{table}

It is straightforward to obtain the expansion of the discontinuity $\Delta(y)$ by termwise analytic continuation $\epsilon\to
e^{i\pi}\epsilon$ in \eqref{kappasingYL},
\begin{eqnarray}\label{DeltaexpYL}
\Delta(y) = (-\epsilon)^\frac{5}{6}\,\tilde K_1(\epsilon)+(-\epsilon)^\frac{5}{3}\,\tilde K_2(\epsilon) + (-\epsilon)^{\frac{5}{2}} \, \tilde K_3(\epsilon) + \cdots
\end{eqnarray}
where $\epsilon = y+Y_0$ is now negative ($\Delta(y)$ vanishes at $y<-Y_0$), and
\begin{eqnarray}\label{Deltaexpcontinuation}
{\tilde K}_n(\epsilon) = \sin(\nu_n\,\pi)\,K_n(\epsilon)\,,
\end{eqnarray}
with $\nu_n$ standing for the exponent in $\epsilon^\nu\,K_n(\epsilon)$ term in \eqref{kappasingYL}. Numerical values of the
coefficients ${\tilde K}_n^{(l)}=(-)^l\,\sin(\pi\nu_n)\,K_n^{(l)}$ at nine leading terms of the expansion \eqref{DeltaexpYL} obtained
from the known coefficients in \eqref{Mepsilon} and \eqref{alphaepsilon} are collected in Tab.\ref{9KappaSingularCoefficients}
\begin{table}[htp]
\begin{center}
\begin{tabular}{|c|c|c|c|c|c|c|c|c|}
\hline
$\tilde K_1^{(0)}$ & $\tilde K_2^{(0)}$ & $\tilde K_1^{(1)}$ & $\tilde K_3^{(0)}$ & $\tilde K_2^{(1)}$ & $\tilde K_1^{(2)}$ & $\tilde K_4^{(0)}$ & $\tilde K_5^{(0)}$ & $\tilde K_3^{(1)}$ \\
\hline
$2.7481$ & $7.274$ & $-1.103$ & $10.78$ & $-1.400$ & $-1.655$ & $-0.5601$ & $10.47$ & $-0.8974$ \\
\hline
\end{tabular}
\end{center}
\caption{Numerical values of the coefficients in the expansion \eqref{DeltaexpYL} with ${\tilde K}_n(\epsilon)=
 \sum_{l=0}^\infty \,{\tilde K}_n^l\,\epsilon^l$.}
\label{9KappaSingularCoefficients}
\end{table}

\begin{figure}[!htp]
\centering
\includegraphics[width=0.85\textwidth]{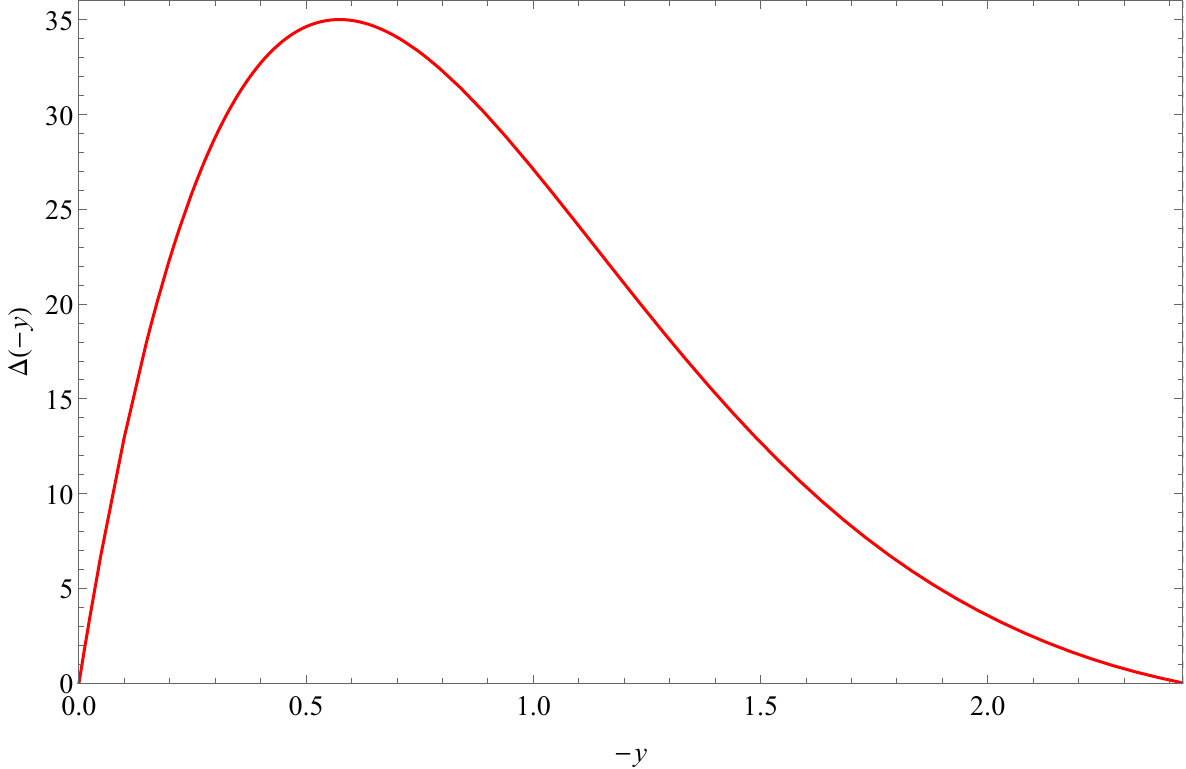}
\caption{Approximation \eqref{Deltaapprox} for the discontinuity $\Delta(-y)$ in the dispersion relation
\eqref{dispeta}.}
\label{bDiscInterpolationCurvature1}
\end{figure}

We propose the following approximation for $\Delta(y)$ in the whole domain of integration in \eqref{dispeta}:
\begin{gather}
\Delta_\text{approx}(y) = \epsilon^{\frac{5}{6}}\tilde K_1^{(0)} + \epsilon^{\frac{5}{3}}\tilde K_2^{(0)} + \epsilon^{\frac{11}{6}}\tilde K_1^{(1)}+ \epsilon^{\frac{5}{2}}\tilde K_3^{(0)} + \epsilon^{\frac{8}{3}}\tilde K_2^{(1)} + \epsilon^{\frac{17}{6}}\tilde K_1^{(2)} \nonumber\\
 + \epsilon^{\frac{19}{6}}\tilde K_4^{(0)} + \epsilon^{\frac{10}{3}}\tilde K_5^{(0)} + \epsilon^{\frac{7}{2}}\tilde K_3^{(1)} + \epsilon^{\frac{11}{3}}\tilde K_2^{(2)} + \epsilon^{\frac{23}{6}}\tilde K_1^{(3)} + \epsilon^{\frac{25}{6}}\tilde K_6^{(0)} \,, \label{Deltaapprox}
\end{gather}
with ${\tilde K}_1^{(0)}, ..., K_3^{(1)}$ taken from Tab.\ref{9KappaSingularCoefficients}, and the last three coefficients
$\tilde K_2^{(2)}\,, \tilde K_1^{(3)} \,,$ and $\tilde K_6^{(0)}$ determined by demanding that $\Delta_\text{approx}(y)$ meets the two conditions \eqref{IntersectionSlope}, and that the dispersion integral \eqref{dispeta} evaluated with $y=-Y_0$ returns returns exact value ${\hat\kappa}(-Y_0)=3$ (see \eqref{kappaYL}),
\begin{equation}
\tilde K_2^{(2)}= -182.959 \,, \quad \tilde K_1^{(3)} = +238.478 \,, \quad \tilde K_6^{(0)} = -66.657 \,.
\end{equation}
The shape of the function $\Delta_\text{approx}(y)$ at $y\in[-Y_0, 0]$ is shown in Fig.\ref{bDiscInterpolationCurvature1}. Despite somewhat artificial way of constructing this approximation, we believe that it is very close to actual discontinuity \eqref{Deltadef}. This expectation can be verified against exact numbers (Sec.2) and numerical data obtained in Sec.5.

\subsection*{Approximation $\boldsymbol{\Delta_\text{approx}(y)}$ vs Data}

In constructing the approximation \eqref{Deltaapprox} we have used the exact slope $\kappa_1$, Eq.\eqref{b1}. However,
the leading term $\kappa_0 = \kappa(\infty)$ in \eqref{kappaexpansioneta} does not contribute to the discontinuity. The dispersion
integral \eqref{dispeta} with our approximation \eqref{Deltaapprox}, evaluated at $\eta=0$, results in
\begin{eqnarray}
\kappa_\text{disp}(\infty) = -\frac{15}{4\pi}\,\int_0^{Y_0}\,\frac{\Delta_\text{approx}(-y)}{y}\,dy = -103.206\,,
\end{eqnarray}
to be compared with the exact value \eqref{binfinity}. The approximation \eqref{Deltaapprox} suggests estimates
for further coefficients in the expansion \eqref{kappaexpu}, e.g.
\begin{eqnarray}
\kappa_{2\,\text{disp}} = -213.278\,, \qquad \kappa_{3\,\text{disp}} = -192.962\,.
\end{eqnarray}

Another exactly known quantity is the slope \eqref{kappauslope} of $\kappa(u)$ at $u=0$. The dispersion relation yields
\begin{eqnarray}
\kappa^{(1)}_\text{disp} = -\frac{15}{4\pi}\,\int_0^\infty\,y^\frac{11}{4}\,\Delta_\text{approx}(-y)\,dy = -56.9019\,.
\end{eqnarray}
Again, it agrees well with the exact value \eqref{kappauslope}. Our approximation gives predictions for several higher terms of
the expansion \eqref{kappaexpu},
\begin{eqnarray}
\kappa^{(n)}_\text{disp} = (-)^n\,\frac{15}{4\pi}\,\int_0^\infty\,y^{\frac{15}{4}\,n-1}\,\Delta_\text{approx}(-y)\,dy\,.
\end{eqnarray}
Thus
\begin{eqnarray}
\kappa_\text{disp}^{(2)} = 317.754\,, \qquad \kappa_\text{disp}^{(3)} = -3524.51\,.
\end{eqnarray}
\begin{figure}[!htp]
\centering
\includegraphics[width=0.75\textwidth]{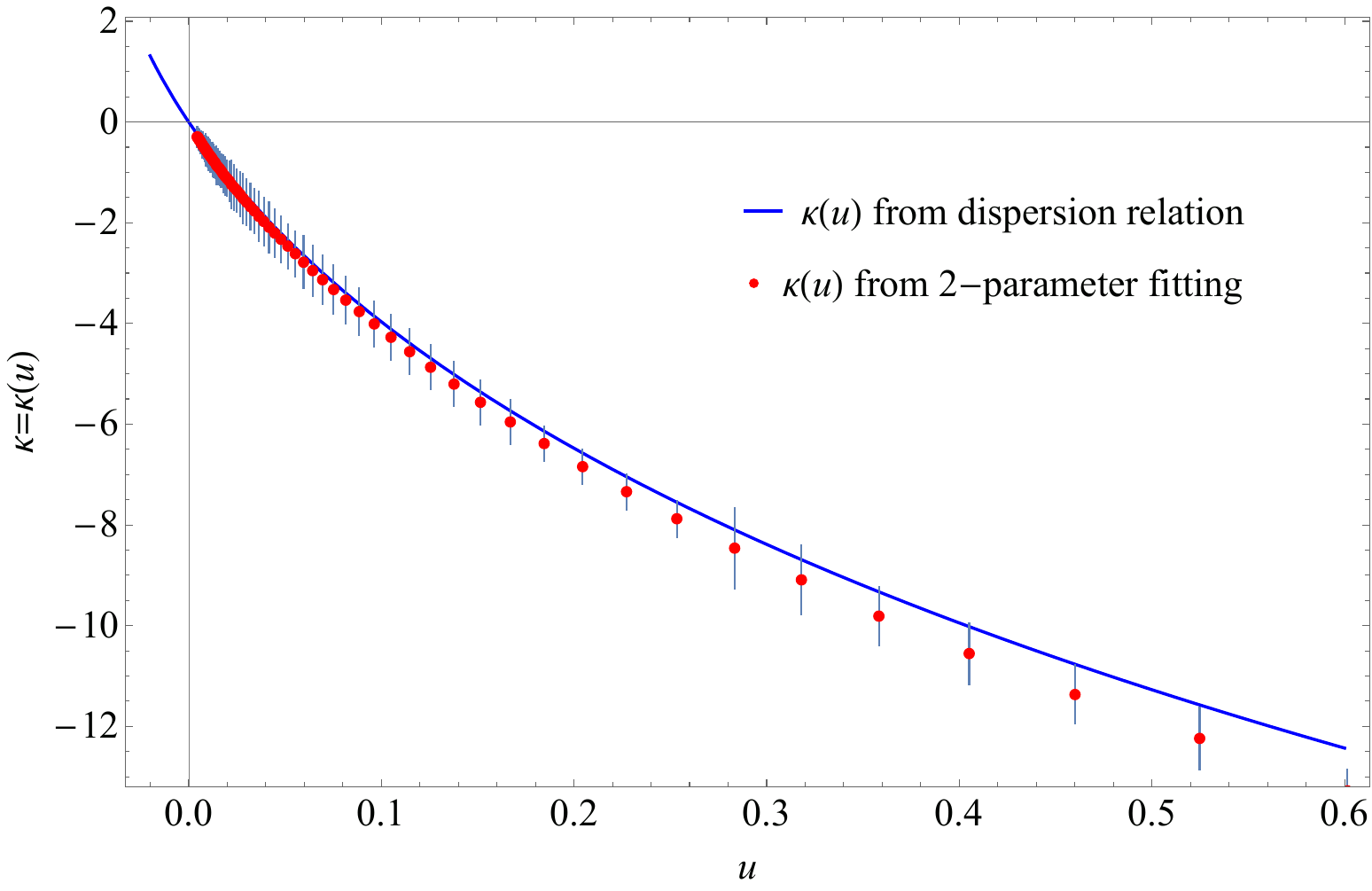}
\caption{Numerical estimates of $\kappa(u)$ obtained by fitting \eqref{kappasigma} to TFFSA data (red bullets), compared to $\kappa_\text{disp}(u)$ from the dispersion relation \eqref{dispeta} with the approximation \eqref{Deltaapprox}.}
\label{KappaFittingPositiveUplot}
\end{figure}

\begin{figure}[!htp]
\centering
\includegraphics[width=0.85\textwidth]{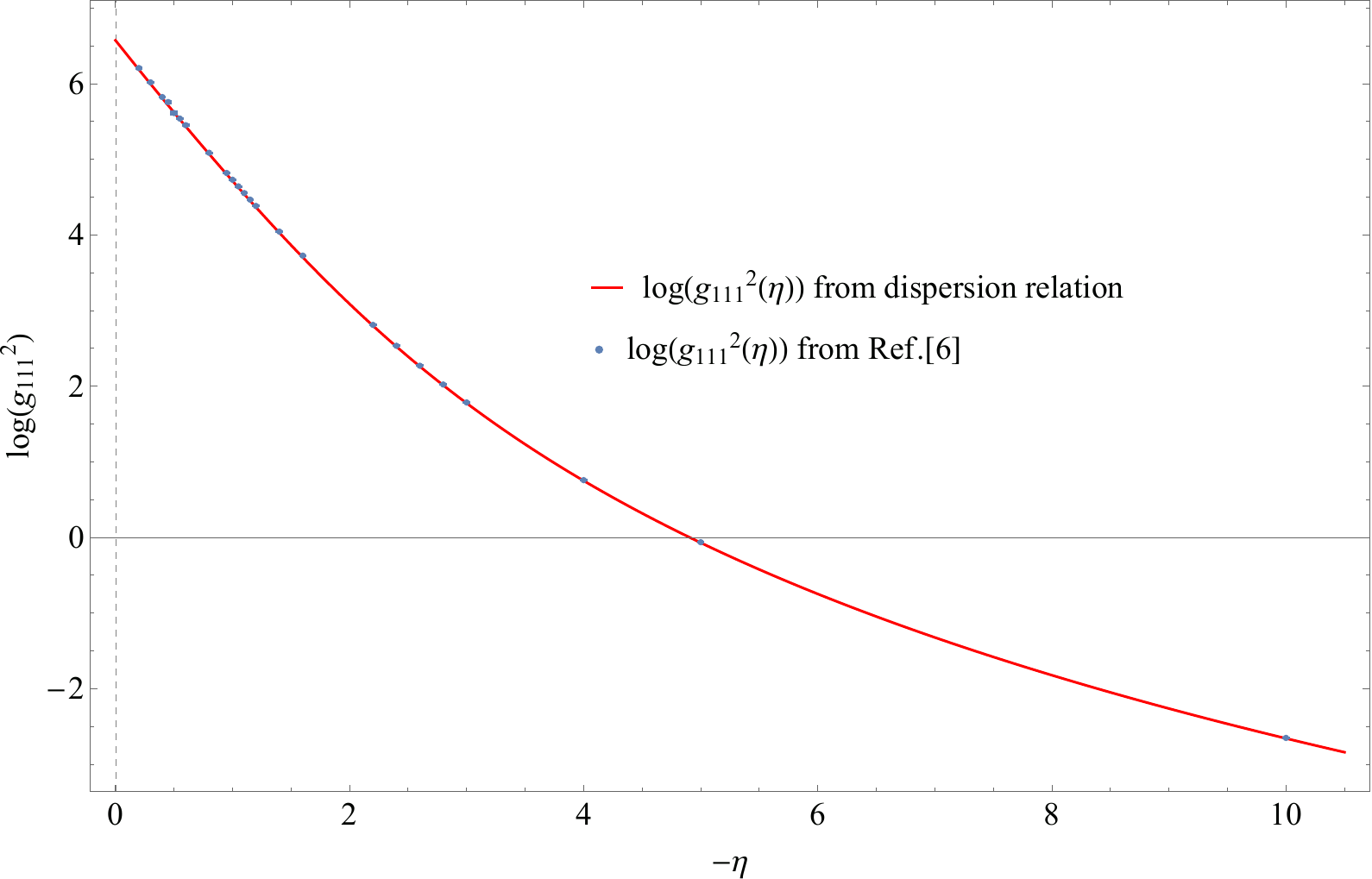}
\caption{Numerical data for $g^2_{111}(\eta):=-4\sqrt{3}\,{\hat \kappa}(\eta)$ from \cite{Gabai:2019ryw} (blue dots), compared with the same quantity evaluated from the dispersion relation \eqref{dispeta} with the approximation \eqref{Deltaapprox}
(red curve). For better visibility, we plot $\log(g_{111}^2)$ against $-\eta$.}
\label{logg111etaDispCheck}
\end{figure}

In the next Section we will use TFFSA to develop numerics for $\kappa(u)$, both at positive and negative $u$. Although
our numerical estimates are not very precise (we present some details of our numerical analysis in Sec.5 below),  Fig.\ref{KappaFittingPositiveUplot} we compare them with $\kappa_\text{disp}(u)$, the integral \eqref{dispeta} evaluated with $\Delta_\text{approx}(y)$ in the integrand. The numerics for $g^2_{111}(\eta):=-4\sqrt{3}\,\hat\kappa(\eta)$ at negative $\eta$ (positive $u$) was previously obtained in \cite{Gabai:2019ryw} using much different analysis of the TFFSA data. We compare these results with $\hat\kappa_\text{disp}(\eta)$ in Fig.\ref{logg111etaDispCheck}. The agreement seems to support the standard analyticity conjecture.

\section{Numerical Data from TFFSA}
Additional data on $\kappa(u)$ can be obtained numerically, using the TFFSA. As was mentioned, TFFSA was applied previously
to estimate the scattering amplitude and in particular the residue \eqref{kappadef} in \cite{Gabai:2019ryw}. We use different approach,
based on the finite-size correction to the two lowest energy levels. TFFSA is a routine which computes the energy levels $E_n(R)$
of the IFT in the geometry of a cylinder, with the spatial coordinate $x$ compactified on a circle of circumference $R$. With sufficiently high truncation level $L$ TFFSA returns very accurate estimates for $E_n(R)$ as long as $R\,\lesssim\, 7 \,|h|^{-8/15}$, but at larger $R$ the accuracy quickly deteriorates due to the truncation effects. All numerics quoted below was obtained at the truncation level $L=13$ \footnote{Here we assume the same definition of the "truncation level" as given in \cite{fonseca2003ising,Xu:2022mmw}; it differs by a factor of $1/2$ from the definition in \cite{Gabai:2019ryw}.}.

\begin{figure}[!h]
\centering
\includegraphics[width=0.5\textwidth]{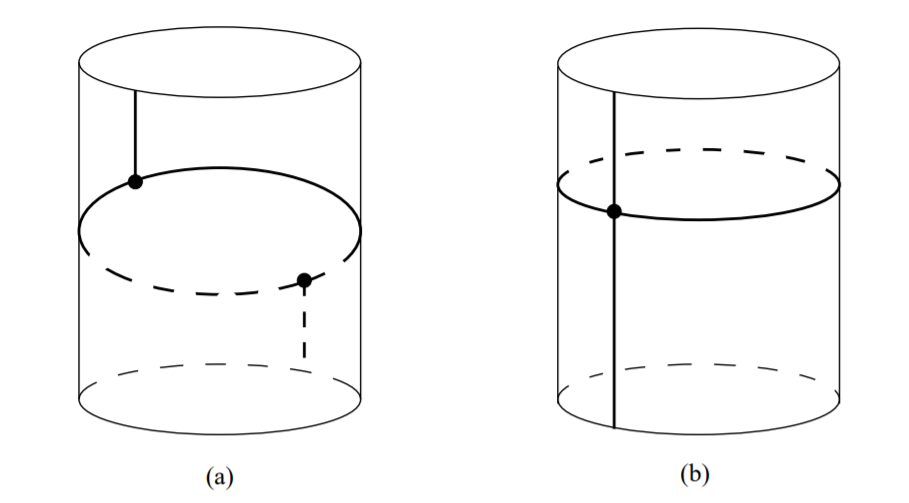}
\caption{The diagrams are presenting the leading finite size corrections to $\Delta E_1(R)$, see \eqref{MassGapExpansion1} and \eqref{MassGapExpansion2}.}
\label{finitesize1loopcorrections}
\end{figure}

Since in this geometry the lowest excited state $n=1$ may be interpreted as a single particle $A_1$ at rest on the circle, the gap
$\Delta E_1(R) :=E_1(R)-E_0(R)$ is expected to converge to $M_1$ as $R\to\infty$. Moreover, when $M_1 \neq 0$ the leading finite-size corrections are expressed exactly in terms of the residue \eqref{kappadef} and the analytic continuation of the amplitude $S(\theta)$ \cite{Yurov:1989yu},
\begin{eqnarray}
\frac{\Delta E_1(R) - M_1}{M_1} = \kappa\, e^{-\frac{\sqrt{3}}{2} M_1 R} - \int \frac{d\theta}{2\pi} \cosh \theta\left[ S\big(\theta+\frac{\pi i}{2}\big) - 1 \right] e^{-M_1 R\cosh \theta} + \cdots \,, \,\, \label{MassGapExpansion1}
\end{eqnarray}
where the dots represent further corrections which decay yet faster at large $R$. The two terms in the r.h.s. of \eqref{MassGapExpansion1} express contributions of the particle $A_1$ winding around the space-time cylinder once, as shown in Fig.\ref{finitesize1loopcorrections}. Other stable particles, when present (as in IFT at $u > u_2 = 0.0640(25)$), lead to additional contributions to the gap
\eqref{MassGapExpansion1}. We describe some of these contributions in Appendix B.

\begin{figure}[!htp]
\centering
\includegraphics[width=0.85\textwidth]{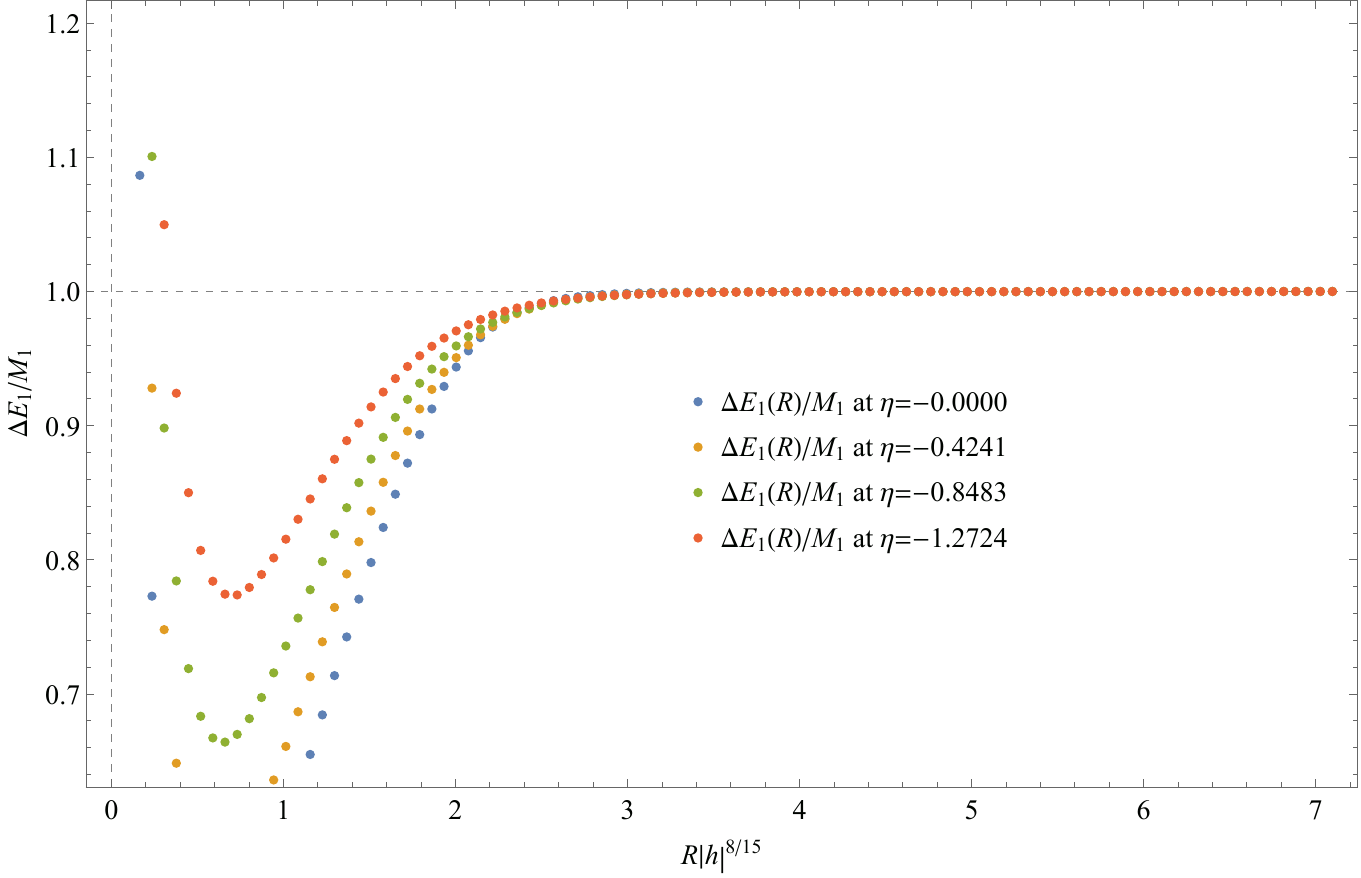}
\caption{Plots of $\Delta E_1(R)/M_1$ at different $\eta$ (real $h$). The exponential approach to $1$ is in accord with \eqref{MassGapExpansion1} with negative $\kappa$.}
\label{DeltaEMrealh}
\end{figure}

\begin{figure}[!htp]
\centering
\includegraphics[width=0.85\textwidth]{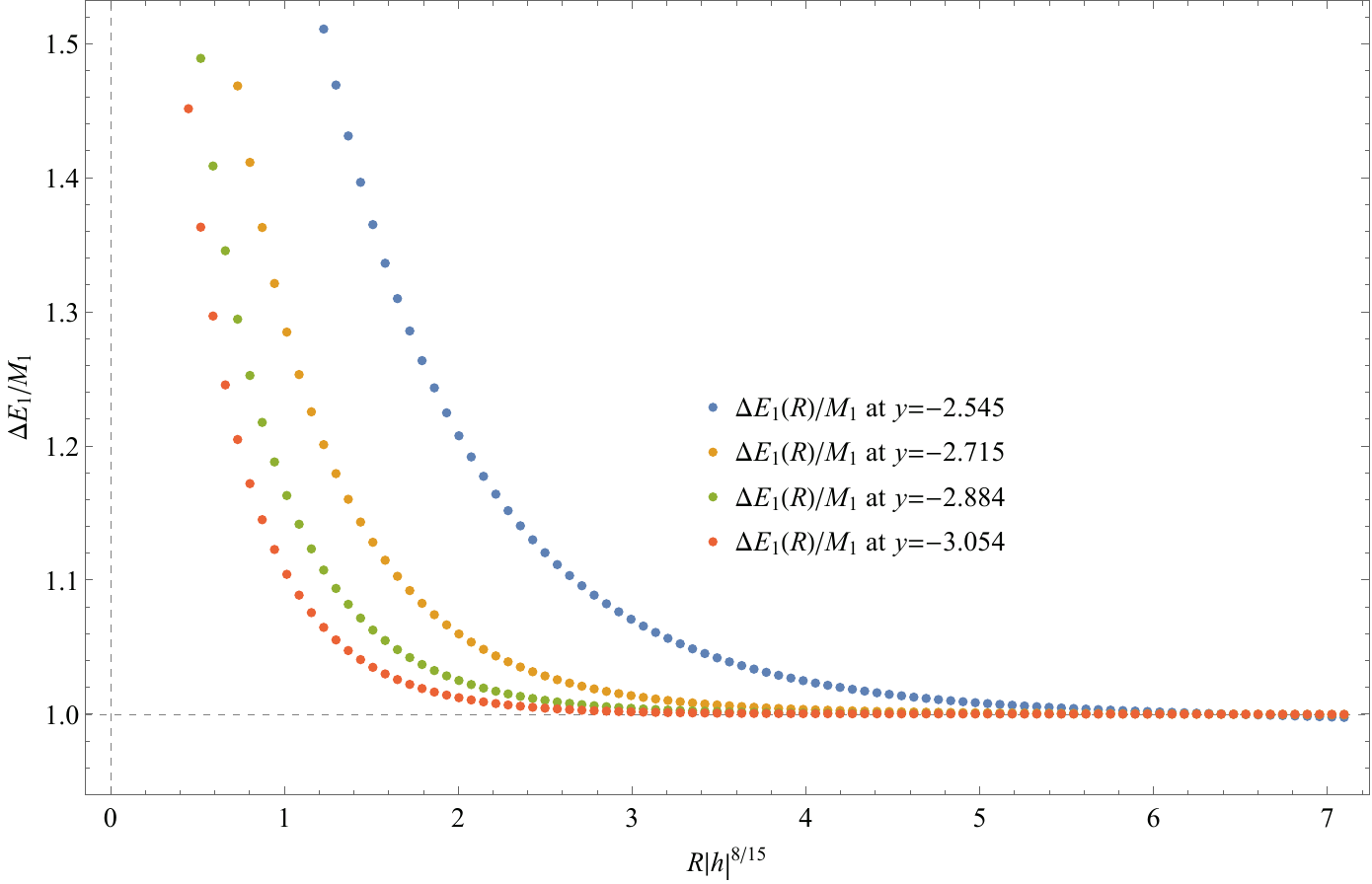}
\caption{Plots of $\Delta E_1(R)/M_1$ at different $y$ for (imaginary $h$). The approach to $1$ is consistent with \eqref{MassGapExpansion1} with positive $\kappa$.}
\label{DeltaEMimaginaryh}
\end{figure}

The gaps $\Delta E_1(R)$ obtained via TFFSA with $L=13$, at few sample values of $u$ are shown in Fig.\ref{DeltaEMrealh} and Fig.\ref{DeltaEMimaginaryh}. The exponential approach to $M_1$ at large $R$ consistent with the leading term in \eqref{MassGapExpansion1} is clearly visible, except for when $u$ is close to the YL point $-u_0$ where the behavior at $R \lesssim 7\,|h|^{-8/15}$ becomes rather power-like. Rough estimate of numerical values of $\kappa(u)$ can be obtained by fitting the leading term in the r.h.s. of \eqref{MassGapExpansion1} to the TFFSA data.

Unfortunately, precision of such estimate is low. The reason is that truncation effects spoil the TFFSA data for $\Delta E_1(R)$ at large $R$ (typically at $MR \gtrsim 20$) where the leading term in \eqref{MassGapExpansion1}
dominates. On the other hand, at smaller values of $MR$ the correction terms in the large $R$ expansion \eqref{MassGapExpansion1} become comparable to the leading exponential term. Thus, the second term in the r.h.s. of \eqref{MassGapExpansion1} has the leading large-$R$ behavior
\begin{eqnarray}\label{secondterm}
-{\sigma}\,(M_1 R)^{-1/2}\,e^{-M_1 R}\,,
\end{eqnarray}
where $\sigma=[S(i\pi/2)-1]/\sqrt{2\pi}$; with $\sigma \sim \kappa$ \footnote{The ratio $\sigma/\kappa$ is $\approx -3.2$ at $u=+\infty$. It becomes positive at $-0.0036 \lessapprox u \lessapprox 12.5$ (turning to zero at the boundaries of this segment) and it approaches $-1.7166..$ when $u\to -u_0$. Obviously, this ratio diverges at $u=0$, where $\sigma$ remains finite; in some vicinity of this point the ration actually is very large, which makes it difficult to estimate $\kappa$ near the point $u=0$.} the first term
dominates only at $MR \gtrsim 10$. Therefore the window in which the above fitting is meaningful is not sufficiently wide, while the fitting results were rather sensitive to the choice of the domain of $R$ where fitting was performed. We estimated the accuracy of this approach by the dependence on the fitting interval to be typically about $15-25\%$. It can be somewhat improved by fitting the data to the two-parameter expression
\begin{eqnarray}\label{kappasigma}
\kappa\, e^{-\frac{\sqrt{3}}{2} M_1 R} - {\sigma}\,(M_1 R)^{-1/2}\,e^{-M_1 R}\,,
\end{eqnarray}
via $\kappa$ and $\sigma$. The data presented in Fig.\ref{KappaFittingPositiveUplot} were obtained through this two-parameter fitting. We estimate the accuracy at generic $u$ as $10-20\%$.

Accuracy of the estimates obtained using \eqref{kappasigma} deteriorates even more when $u$ is close to one of the three special points mentioned in Sec.3. Since the correlation length $M^{-1}$ diverges at the YL point, the window available for the fitting shrinks as $u$ approaches $-u_0$. When $u$ is close to zero $\kappa(u) \sim u$ becomes small, and within the acceptable range
it is barely detectable on the background of the higher finite-size corrections. In vicinity of the $E_8$ point $\eta=0$ the theory involves more than one stable particle; these additional particles give rise to additional sub-leading terms which are significant in the interval of $R$ available for fitting. However, near these points the accuracy can be significantly improved with the use of exact
scattering amplitudes at these points.

\subsection*{Vicinity of Free Fermion point}

\begin{figure}[!htp]
\centering
\includegraphics[width=0.75\textwidth]{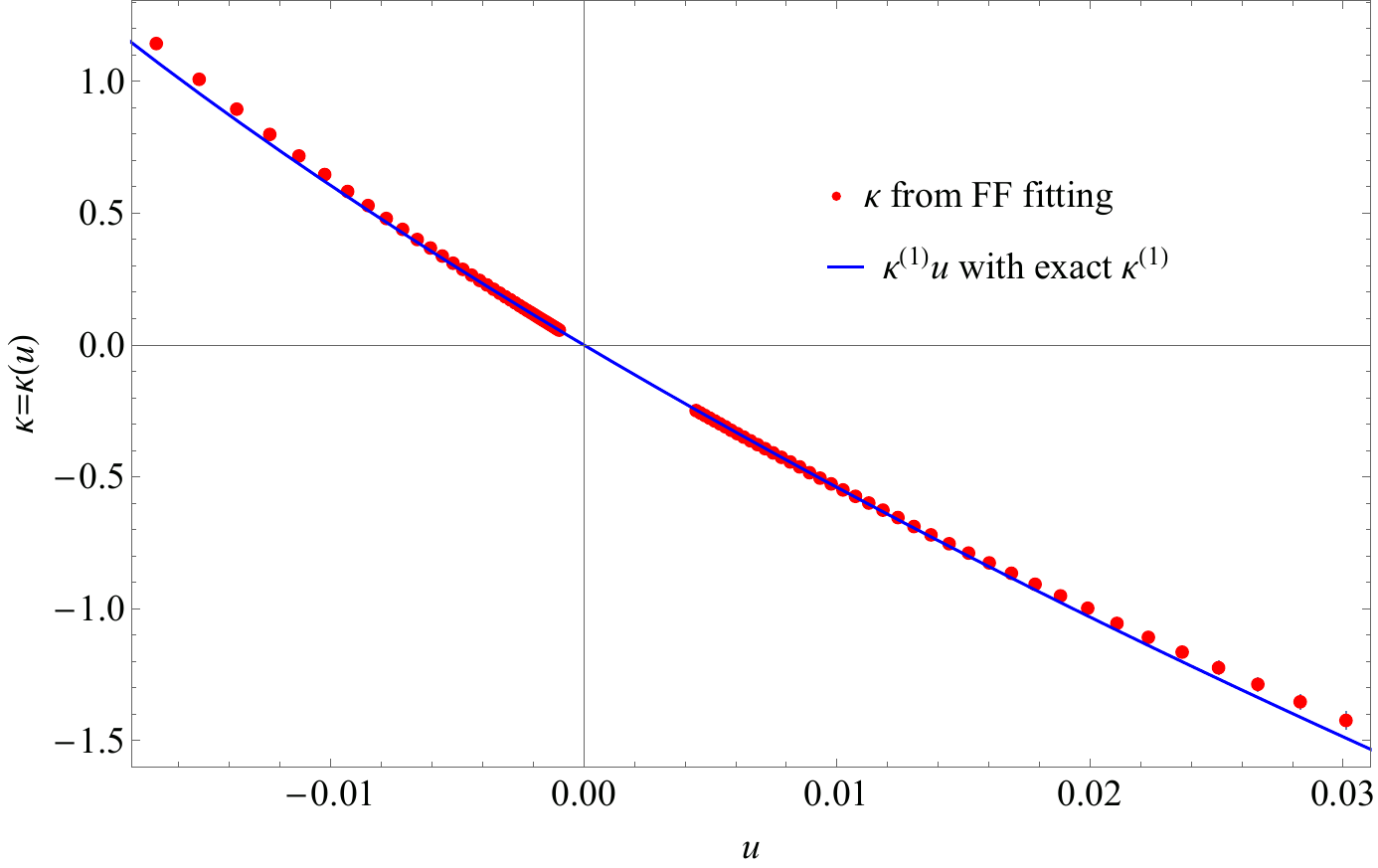}
\caption{The results of the fitting \eqref{DelEkappaFF} to TFFSA data via $\kappa$, at $-0.02<u<+0.03$, are shown as red bullets. The blue curve is $\kappa_\text{disp}(u)$, the integral \eqref{dispeta} with $\Delta(y)$ given by \eqref{Deltaapprox}.}
\label{KappaFittingNearFFplot}
\end{figure}

At $u=0$ we are dealing with the free fermion theory, with trivial S-matrix $S(\theta) = -1$ and $\kappa(0) = 0$. In high-T domain, the finite size ground state is the Neveu-Schwarz vacuum, while the first excited state is understood as one particle $A_1$ at rest
over the Ramond vacuum. Therefore, at $u=0$ the finite size gap is given by
\begin{equation}\label{FFgap}
\frac{\Delta E^{\text{FF}}_1(R) - M_1}{M_1} = - \int_{-\infty}^{+\infty} \frac{d\theta}{2\pi} \cosh \theta \, \log \Big( \frac{1-e^{-M_1 R \cosh\theta}}{1+e^{-M_1 R \cosh\theta}} \Big)\,,
\end{equation}
which accounts for contributions from particles $A_1$ winding around the cylinder any number of times. At small $u$ the amplitude $S(\theta)$ acquires a correction
\begin{equation}
S(\theta) = -\Big( 1 + \frac{i \kappa A(\theta)}{\sinh \theta} \Big) \,, \qquad A(\theta) = A_{\text{pole}}(\theta) + A_\sigma(\theta) \,,
\end{equation}
where:
\begin{equation}\label{Apole}
A_\text{pole}(\theta) = -\frac{2}{\sqrt{3}}\,\,\frac{\sinh^2 \theta}{\sinh^2 \theta + \frac{3}{4}}\,,
\end{equation}
and the absorptive part $A_\sigma(\theta)$ is determined by the inelastic processes, see \cite{Zamolodchikov:2011wd}.
This leads to correction terms to the finite size gap \eqref{FFgap},
\begin{eqnarray}
&&\frac{\Delta E_1(R) - M_1}{M_1} = - \int_{-\infty}^{+\infty} \frac{d\theta}{2\pi} \cosh \theta \, \log \Big( \frac{1-e^{-M_1 R \cosh\theta}}{1+e^{-M_1 R \cosh\theta}} \Big) + \qquad \qquad\nonumber\\
&&\qquad\qquad +\kappa \, e^{-\frac{\sqrt{3}}{2}M_1 R}  - \frac{2}{\sqrt{3}} \, \kappa \int\frac{d\theta}{2\pi}
\, A(i\pi/2+\theta)\,e^{-M_1 R \cosh\theta}\,. \label{DelEkappaFF}
\end{eqnarray}
When $M_1 R \gtrsim 1$ the integral in \eqref{DelEkappaFF} is dominated by $\theta \lesssim 1$, where $A_\sigma(\theta)$ is numerically small as compared to $A_\text{pole}(\theta)$ (see Fig.3 in \cite{Zamolodchikov:2011wd}), and one  can neglect its contribution to the finite size gap. Therefore, we have used \eqref{DelEkappaFF} with $A(\theta)=A_\text{pole}(\theta)$ to fit TFFSA data via $\kappa$. The best fit was obtained in the interval $ 3.5 < M_1 R < 9.5$, and the result is shown in  Fig.\ref{KappaFittingNearFFplot}, where it is compared to $\kappa_\text{disp}(u)$ obtained from the dispersion relation \eqref{dispu}.

\subsection*{Vicinity of $E_8$ point}
When negative $\eta$ is sufficiently close to the $E_8$ point $\eta=0$, IFT involves more than one stable particle.
At $\eta_3 < \eta <0$, $\eta_3 \approx -0.13$, there are two heavier particles $A_2$ and $A_3$, and at $\eta_2 <\eta< \eta_3$, $\eta_2 \approx -2.08$, only one additional
stable particle $A_2$ remains in the spectrum\footnote{There are eight stable particles at $\eta=0$, see footnote 2 in Sec.2. When $\eta$ shifted from zero five of them loose stability and become resonance states.}\ \cite{zamolodchikov2013ising}. While the finite-size correction terms displayed in \eqref{MassGapExpansion1} describe contributions of the particle $A_1$ winding once around the cylinder, similar winding of these heavier particles generates further sub-leading terms in the large $R$ expansion of $\Delta E_1 (R)$,
\begin{gather}
\frac{\Delta E_1(R) - M_1}{M_1} = \kappa\, e^{-\frac{\sqrt{3}}{2} M_1 R} + \kappa_2\, e^{- \mu^1_{22} R} + \kappa_3\, e^{-\mu^1_{33} R} -\nonumber\\
\sum_{p=1}^{3} \frac{M_p}{M_1} \int \frac{d\theta}{2\pi} \cosh \theta\big[ S_{1p}(\theta+\frac{\pi i}{2}) - 1 \big] e^{-M_p R\cosh \theta} + O(e^{-2 M_1 R}) \,, \,\, \label{MassGapExpansion3terms}
\end{gather}
where the notations are explained in Appendix B. Of course when $\eta < \eta_3$ the terms associated with the particle $A_3$
(i.e. the exponential term with $\kappa_3$ as well as the term with $p=3$ in the second line in \eqref{MassGapExpansion3terms})
are to be dropped.

\begin{figure}[!htp]
\centering
\includegraphics[width=0.75\textwidth]{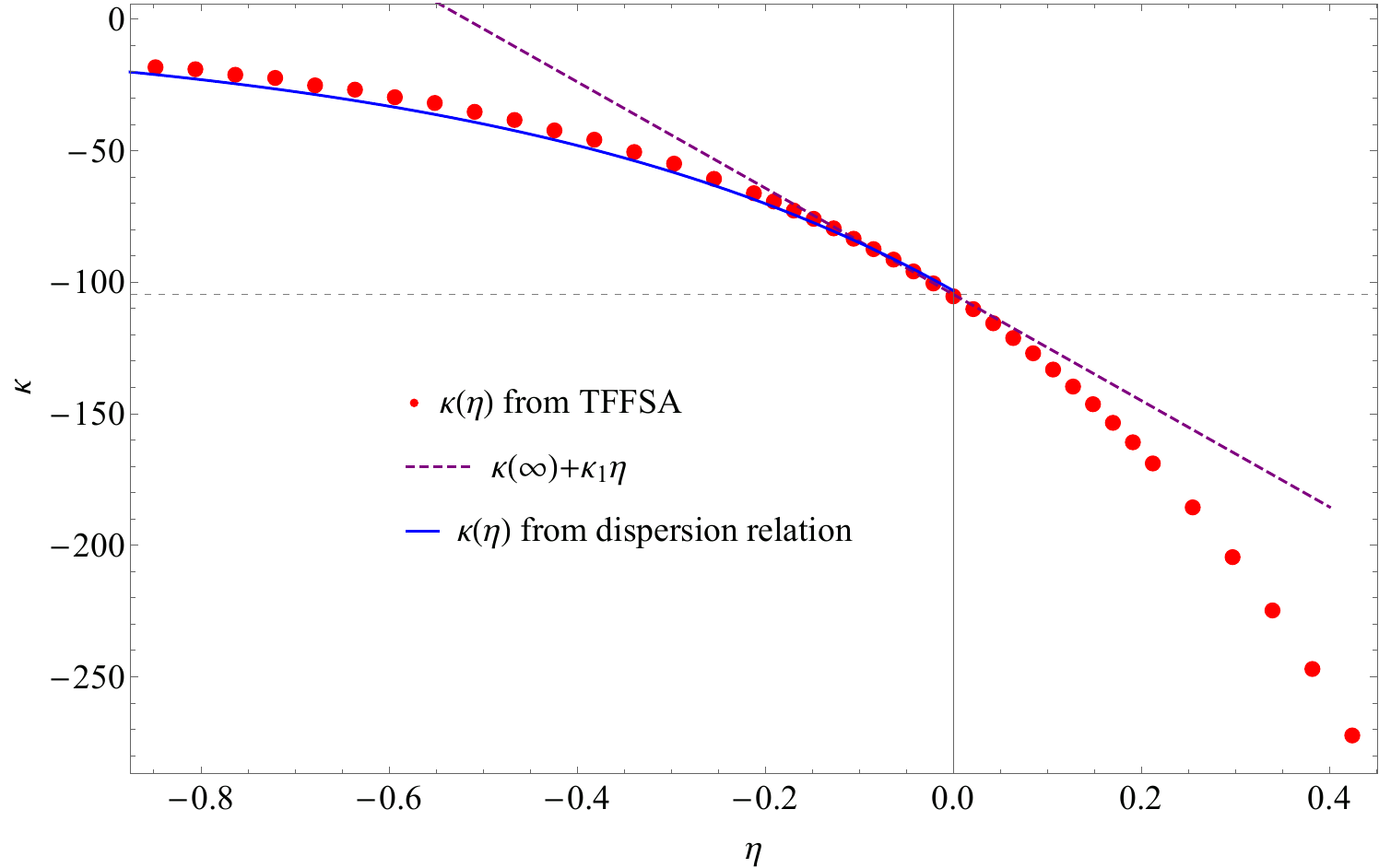}
\caption{Red bullets show the result of the best fit of $\hat \kappa(\eta)$ to TFFSA data for $\Delta E_1(R)$, using \eqref{MassGapExpansion3terms} with $\kappa$ and $\kappa_2$ ($\kappa_3$ was set to zero, see main text). The blue curve is ${\hat\kappa}_{\text{disp}}(\eta)$, the integral \eqref{dispeta} with $\Delta(y)$ given by \eqref{Deltaapprox}, and the magenta dashed line is showing the exact slope $\kappa_1$ at $E_8$ point, see \eqref{b1} and Appendix A.}
\label{realkappafitting}
\end{figure}

We used \eqref{MassGapExpansion3terms} as a fit of the TFFSA data for $\Delta E_1(R)$ at $-0.6 < \eta < 0$. To make \eqref{MassGapExpansion3terms} useful for determining $\kappa$ one needs to know the masses $M_2$ and $M_3$ in this
interval. We have determined $M_2$ and $M_3$ directly from the TFFSA data for $M_1 R \gg 1$. Also, $\kappa_3$ must be set to zero at $\eta<\eta_3$, and
moreover one can argue that the parameter $\kappa_3$ is rather small at $\eta\in [\eta_3 .. 0]$. Indeed, $\kappa_3$ must turn to zero
at the boundaries of this narrow interval. For that reason we set $\kappa_3$ to zero in the whole domain $[-0.6 .. 0]$. Finally,
we assumed that when $\eta$ is small the S-matrix elements $S_{1 p}(\theta)$ are not too much different from these amplitudes at $\eta=0$ which are known exactly. With these approximations, $\kappa$ was determined from the best fit to the
TFFSA via two parameters $\kappa$ and $\kappa_2$. In all cases some segments within the window $12.2<M_1R<18.8$ were taken as the fitting domain. This procedure substantially improved the accuracy of the estimate of $\kappa$, especially when $\eta$ is close to zero ($\eta \gtrsim -0.6$). The results are shown in Fig.\ref{realkappafitting}, where the error bars reflect the dependence on the choice of the fitting interval.

\begin{figure}[!htp]
\centering
\includegraphics[width=0.75\textwidth]{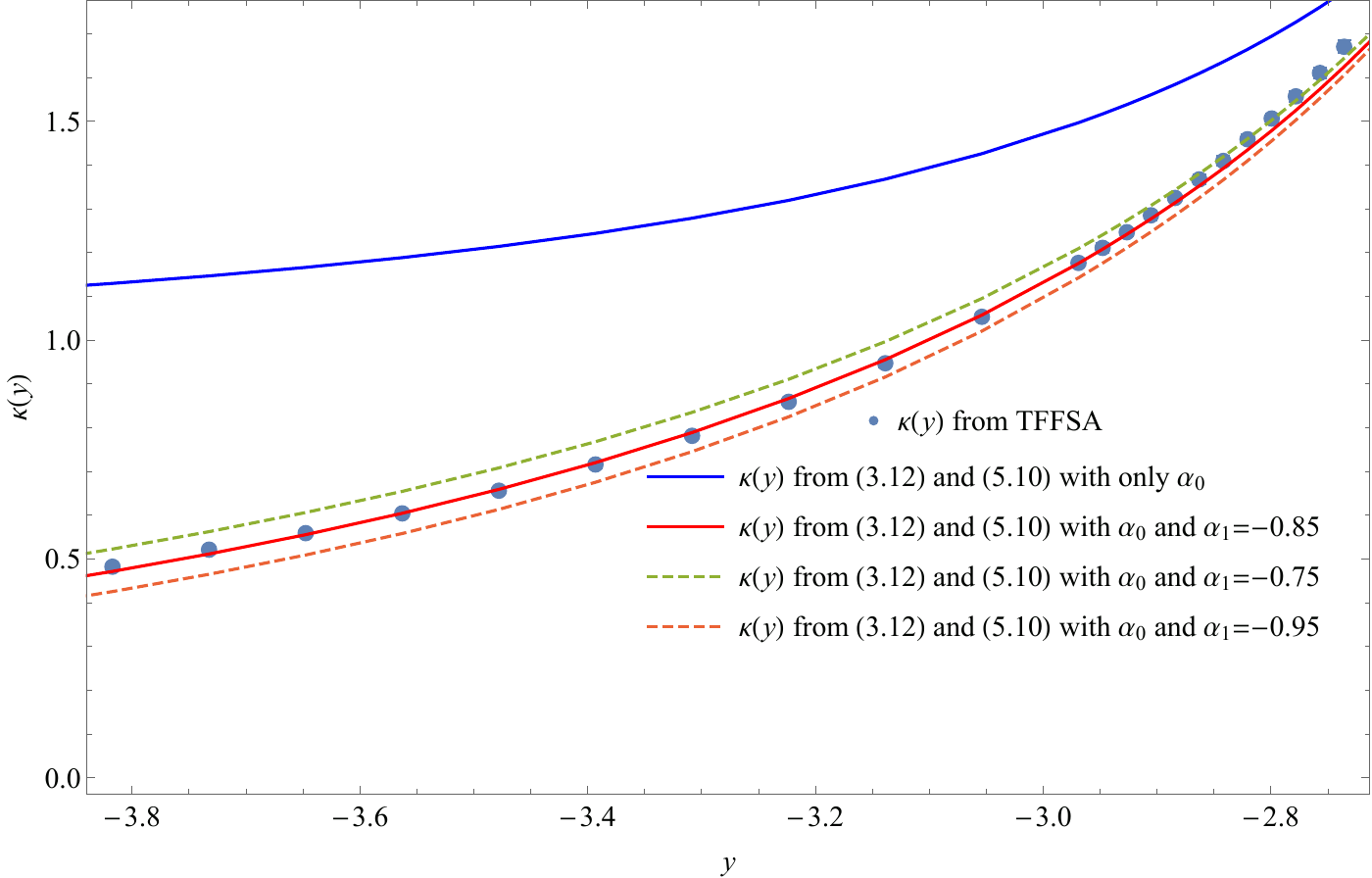}
\caption{Numerical estimates of ${\tilde\kappa}(y):=\kappa((-y)^{-15/4})$ obtained by fitting \eqref{MassGapExpansionYLfitting}
to TFFSA data via $\kappa$ (grey bullets). The blue line shows ${\tilde\kappa}(y)$ given by \eqref{kappattbar} with $\kappa^{(0)} = 3$ and set $\alpha \approx \alpha_0=-1.32\,|m|^{-2}$, respectively. The red line is the plot of \eqref{kappattbar} with $\kappa^{(0)}=3$, and $\alpha$ given by \eqref{alphatwoterms} with ${\hat\alpha}_1 = -0.85\,|m|^{-2}$. The dotted lines
show the same function with slightly different values of ${\hat\alpha}_1$.}
\label{imaginarykappafittingwithalpha1}
\end{figure}

\subsection*{Vicinity of the YL point}

When close to the YL point there is a good approximation for the amplitude $S(i\pi/2 +\theta)$ in the second term
in \eqref{MassGapExpansion1}. The main correction due to the irrelevant operators in \eqref{aeff0} to the YLQFT S-matrix \eqref{SmatrixYL} comes from the operator $(T{\bar T})$ in \eqref{aeff1}. As was explained in Sec.3, up to the order
$\alpha^4 M_1^8 \sim (u+u_0)^\frac{10}{3}$, its contribution is identical to the TTbar deformation \eqref{sttbar}, with
the $S^{(0)}$ taken to be the YLQFT S-matrix \eqref{SmatrixYL}. Therefore near the YL point \eqref{MassGapExpansion1} can be approximated as
\begin{eqnarray}\label{MassGapExpansionYLfitting}
&&\frac{\Delta E_1(R) - M_1}{M_1} = \kappa\, e^{-\frac{\sqrt{3}}{2} M_1 R} - \qquad\qquad\qquad\\
&&\qquad\qquad \int \frac{d\theta}{2\pi} \cosh \theta\left[ e^{-M_1 R\cosh\theta}\,e^{ \alpha M_1^2 \cosh \theta}   \, S_{11}^\text{YLQFT}\big(\theta+\frac{\pi i}{2}\big) - 1 \right] + \cdots \nonumber\,\,
\end{eqnarray}
with the mass $M_1$ given by the expansion \eqref{Mexp}. We also used two term approximation for $\alpha$,
\begin{eqnarray}\label{alphatwoterms}
\alpha = \alpha_0 +\hat\alpha_1\,(y+Y_0)\,,
\end{eqnarray}
with known $\alpha_0 = -1.32$ \cite{Xu:2022mmw}, and adjustable slope $\alpha_1$. With this approximations of the second term in \eqref{MassGapExpansion1} we found that the best fit of \eqref{MassGapExpansion1} to the TFFSA data via $\kappa$ can be obtained with
\begin{eqnarray}
\hat\alpha_1 |m|^2 = -0.85 \pm 0.1\,.
\end{eqnarray}
When the fitting interval was taken within the segment $M_1 R \in [4.0,7.5]$ the fitting results were stable for $-3.8 < y < -2.8$. As we approached yet closer to the YL point $-Y_0 = -2.4293$ the quality of the fit deteriorated rapidly. The mass $M_1 \simeq 1.602\ \epsilon^{5/6}$ becomes small, and the domain of $R$ where \eqref{MassGapExpansionYLfitting} is expected to give meaningful approximation is pushed beyond the range where TFFSA data is accurate. The result is shown in Fig.\ref{imaginarykappafittingwithalpha1}, along with ${\tilde\kappa}(y):=\kappa((-y)^{-15/4})$ from the approximation \eqref{kappattbar} with $\alpha=\alpha_0$ and $\alpha$ given by \eqref{alphatwoterms}.

\section{Discussion}

In this work we have tested analyticity of the $\varphi^3$ coupling $\kappa(u)$ as the function of the scaling parameter
$u = h^2/|m|^\frac{15}{4}$. Assuming maximal analyticity, we have constructed an approximation for the discontinuity across the branch cut in Fig.1, and compared the dispersion relation \eqref{dispu} with exact and numerical data available through TFFSA in Sec.5, and previously in \cite{Gabai:2019ryw}. Agreement supports the analyticity conjecture, as well as the approximation \eqref{Deltaapprox} for the discontinuity.

As in \cite{Xu:2022mmw}, important tool in our analysis was application of the $T{\bar T}$ deformation formulae, which allowed
for control of higher order terms in the coupling parameter $\alpha(u)$ in the effective action \eqref{aeff1}. For $\kappa(u)$
the equation \eqref{kappattbar} reproduces exactly the $\alpha$-expansion up to terms $\alpha^4$, leading to the singular
expansion \eqref{kappaexp3}, \eqref{kappa0exp}. The terms $\sim \alpha^5$ and higher interfere with the contributions from the
operators $(T{\bar T})^3 = L_{-2}^3 {\bar L}_{-2}^3 I$ and higher descendants of identity. We would like to note in this connection that the operator $(T{\bar T})^3$, as well as $(T{\bar T})^4:=L_{-2}^4{\bar L}_{-2}^4 I$ and a string of certain higher descendants of identity, generate the "generalized TTbar deformation" \cite{smirnov2017space} (see also \cite{Camilo:2021gro}) with known effect on the S-matrix. This gives access to yet higher
terms in the singular expansion \eqref{kappaexp1}, and potentially may throw some light on important general question of convergence
of the effective action expansion \eqref{aeff0}: Does it (or, rather the associated singular expansions like \eqref{kappasing0})
converge in some finite domain around the YL point? We hope to address this intriguing problem in the future.

One of practical outcomes of this work is the first estimate \eqref{alpha1y} of the sub-leading term $\alpha_1$ in \eqref{alphaepsilon}, which
would help to further constrain coefficients in singular expansions of physical quantities around the YL point. This is important
in view of potential application to the problem of "extended analyticity" which would include the low-T domain. Unfortunately,
our estimate \eqref{alpha1y} is not too precise, with possible error of about 20\%. Further work is required to improve this
result, and to determine higher couplings in the effective action \eqref{aeff1}.

We regard this work as a step toward quantitative understanding of how the particle spectrum (including resonance states) and S-matrix of IFT depends on the parameters of the theory. Except of some points in the parameter space, the theory is not integrable, and its S-matrix involve all kinds of inelastic processes. We will address some properties of inelastic amplitudes and resonance states in forthcoming work \cite{XuEtAl2022}. Significance of this study is two-fold. First, IFT describes the basic universality class of second order phase transitions (such as the liquid-vapor critical point), and its analysis is important in understanding
detailed structure of the associated critical singularity. On the other hand, we develop tools for quantitative description of
strongly interacting QFT with rich particle/resonance spectrum and S-matrix.

\subsection*{Acknowledgments:}
AZ is grateful to V.Bazhanov and F.Smirnov for interest to this work and very useful discussions, and HLX thanks B.McCoy and R.Shrock for helpful discussions and comments. Research of AZ was supported in part by National Science Foundation under Grant PHY-1915093 and PHY-2210533.\\

\noindent\textbf{Note added:} In forthcoming paper \cite{BazhanovYL} the free energy of IFT in imaginary magnetic field was computed numerically using an improved corner transfer matrix approach. The authors managed to locate the YL singularity at $u_0=0.0358536516(1)$, with the precision which seems to be substantially better then our estimate $u_0 = 0.03586(4)$. In addition, they give what appears to be more precise estimate of some of the parameters in the effective action \eqref{aeff1}, particularly the coupling $\alpha_0$. We are grateful to the authors of \cite{BazhanovYL} for sharing these results prior to publication.

\appendix

\section*{Appendix}

\section{Derivation of $\kappa_1$ in {\eqref{b1}}}

Consider again the $A_1 + A_1 \to A_1 + A_1$ elastic scattering amplitude $S_{11}(\theta)$. This amplitude of course depends
on the scaling parameter $\eta$, but we suppress this argument to simplify notations. Analytic continuation of this amplitude to complex $\theta$ yields the function analytic in the $\theta$-plane with the poles manifesting the bound states, and the branch cuts associated with the inelastic channels $A_1 + A_1 \ \to X$, where $X$ stands for any states involving more than two of the particles $A_1$, as well as the higher bound states $A_p$. The principal branch values of this amplitude enjoy the functional relations
\begin{eqnarray}\label{S11crossunitarity}
S_{11}(\theta)S_{11}(-\theta) =1\,,\qquad S_{11}(\theta) = S_{11}(i\pi-\theta)\,,
\end{eqnarray}
which follow from unitarity and analyticity (see e.g. \cite{Zamolodchikov:2011wd} for the details).

At $\eta=0$ the theory is integrable. We denote $S_{11}^{(0)}(\theta)$ the amplitude at $\eta=0$. At this integrable point
all inelastic channels are closed, and $S_{11}^{(0)}(\theta)$ is a meromorphic function of $\theta$. Its exact form is
\begin{eqnarray}\label{S11-E8}
S_{11}^{(0)} (\theta):=\frac{\sinh\theta+i\sin\frac{2\pi}{3}}{\sinh\theta-i\sin\frac{2\pi}{3}}\,\frac{\sinh\theta+i\sin\frac{2\pi}{5}}
{\sinh\theta-i\sin\frac{2\pi}{5}}\,\frac{\sinh\theta+i\sin\frac{\pi}{15}}{\sinh\theta-i\sin\frac{\pi}{15}}\,,
\end{eqnarray}
The poles at $\theta=2\pi i/3$, $\theta=2\pi i/5$, and $\theta= i\pi/15$ are the manifestations of the particles $A_1$, $A_2$, and $A_3$ in the direct channel of the scattering, while the poles at $\theta=i\pi/3$, $\theta=3\pi i/5$, and $\theta=14\pi i/15$ represent the same particles in the cross channel\footnote{Although the theory at this integrable point involves eight stable particles, only three lightest particles $A_1$, $A_2$, and $A_3$ appear as the poles in $S_{11}^{(0)}(\theta)$. The remaining particles show up in as the poles in the elastic amplitudes of $S_{p p'}^{(0)}(\theta)$ with $p$ or $p'$ greater than one, see \cite{zamolodchikov1989integrals} for details.}. Correspondingly, the masses of the particles $A_2$ and $A_3$ at $\eta=0$ are
\begin{eqnarray}
M_2^{(0)} =2M_1^{(0)}\,\cos\left(\frac{\pi}{5}\right)\,, \qquad M_2^{(0)} = 2 M_1^{(0)}\,\cos\left(\frac{\pi}{30}\right)\,,
\end{eqnarray}
where $M_1^{(0)}$ is the mass of $A_1$ at $\eta=0$ \footnote{Closed form expression for the coefficient $M_1^{(0)}/|h|^\frac{8}{15}$ can be found in e.g. \cite{Xu:2022mmw}, Appendix B.}.

At nonzero but sufficiently small $\eta$ the elastic amplitude $S_{11}(\theta)$ admits convergent expansion in powers of $\eta$,
which we choose to write as
\begin{eqnarray}\label{S11expansion}
S_{11}(\theta)= S_{11}^{(0)}(\theta)\,\left(1 + \eta\,\,\Phi^{(1)}(\theta) + \frac{1}{2}\eta^2\,\,\Phi^{(2)}(\theta) + ...\right)\,.
\end{eqnarray}
Here we concentrate attention on the first correction term $\eta\,\varphi^{(1)}(\theta)$, which satisfy the relations
\begin{eqnarray}
\Phi^{(1)}(\theta) = -\Phi^{(1)}(-\theta)\,, \qquad \Phi^{(1)}(\theta)=\Phi^{(1)}(i\pi-\theta)\,,
\end{eqnarray}
in virtue of \eqref{S11crossunitarity}. In fact, $\Phi^{(1)}(\theta)$ can be found in a closed form, as follows.

At nonzero $\eta$ the integrability of IFT is broken, and $S_{11}(\theta)$ develops branching points at the
inelastic thresholds. However, the discontinuities of this amplitude across the associated branch
cuts are $O(\eta^2)$, and hence $\varphi^{(1)}(\theta)$ in \eqref{S11expansion} remains a meromorphic function of $\theta$.
Indeed, the discontinuities across the branch cuts are related in a simple way to the inelastic cross sections, which in turn
are proportional to the squares of the absolute values of the inelastic amplitudes. Since at small $\eta$ the inelastic amplitudes
are $O(\eta)$, the discontinuities are $O(\eta^2)$. It is then possible to argue that, up to terms $\eta^2$ and higher, the
amplitude \eqref{S11expansion} can be written as
\begin{eqnarray}\label{S11form}
S_{11}(\theta) = \frac{\sinh\theta + i\sin\frac{2\pi}{3}}{\sinh\theta - i\sin\frac{2\pi}{3}}\,\frac{\sinh\theta + i \sin\alpha_2}
{\sinh\theta - i \sin\alpha_2}\,\frac{\sinh\theta + i \sin\alpha_3}{\sinh\theta - i \sin\alpha_3} + O(\eta^2)\,,
\end{eqnarray}
where
\begin{eqnarray}\label{alphaslopes}
\alpha_2 = \frac{2\pi}{5} + \alpha_2'\,\eta\,, \qquad \alpha_3 = \frac{\pi}{15} + \alpha_3'\,\eta\,,
\end{eqnarray}
and $\alpha_p'\,, \ p=2,3$ are numbers, to be determined shortly. In other words, the leading $\sim \eta$ correction
to $S_{11}(\theta)$ is determined entirely by the shifts of the positions of the poles (and related zeros) of
$S_{11}(\theta)$ associated with the particles $A_2$ and $A_3$. Expanding \eqref{S11form} to the first order in $\eta$ one finds
\begin{eqnarray}\label{varphi1}
\Phi^{(1)}(\theta) = \frac{2i\,\alpha_2'\,\cos\frac{2\pi}{5} \, \sinh\theta}{\sinh^2\theta +\sin^2\frac{2\pi}{5}} + \frac{2i\,\alpha_3'\,\cos\frac{\pi}{15} \, \sinh\theta}{\sinh^2\theta +\sin^2\frac{\pi}{15}}\,.
\end{eqnarray}
and hence
\begin{eqnarray}\label{k1A}
\kappa_1 = \sqrt{3}\,\left[\frac{\alpha_2'\,\cos\frac{2\pi}{5}}{\sin^2\frac{2\pi}{3}-\sin^2\frac{2\pi}{5}}+
\frac{\alpha_3'\,\cos\frac{\pi}{15}}{\sin^2\frac{2\pi}{3}-\sin^2\frac{\pi}{15}}\right]\,\kappa(\infty)
\,,
\end{eqnarray}
where $\kappa_0=\kappa(\infty)$, see Eq.\eqref{binfinity}.

The slopes $\alpha_p'$ in \eqref{alphaslopes} are related in a simple way
\begin{eqnarray}\label{alphaslopes}
\alpha_p' = \frac{2}{\sqrt{\big( 2 M_1^{(0)}/M_p^{(0)}\big)^2-1}}\,\left(\frac{M_1^{(1)}}{M_1^{(0)}} -\frac{M_p^{(1)}}{M_p^{(0)}}\right)\,,
\end{eqnarray}
to the leading coefficients in the expansions of the masses $M_p$ of the particles $A_p$,
\begin{eqnarray}\label{Mpexpansions}
M_p = M_p^{(0)} + M_p^{(1)}\,\eta + O(\eta^2)\,.
\end{eqnarray}
While $M_p^{(0)}$ are known exactly from the integrable theory, the leading corrections $M^{(1)}_p$ were
obtained via the form factor perturbation theory in \cite{Delfino:1996xp}, leading to
\begin{eqnarray}
\alpha_2' = 0.378325... \,, \qquad \alpha_3' = 1.35226...\,.
\end{eqnarray}
With this, \eqref{k1A} leads to \eqref{b1}.

One could rise two objections to the proposed expression \eqref{varphi1}. First, \eqref{varphi1} seems to ignore
the possibility of appearance of new poles, not directly related to the particles $A_2$ and $A_3$. It is easy to
argue though that such new poles can appear (and indeed do appear) only in the order $O(\eta^2)$, i.e. in $\Phi^{(k)}(\theta)$
with $k\geq 2$ in \eqref{S11expansion}. Assume that a new pole appears in $\Phi^{(1)}(\theta)$ at
$\theta=\theta_\text{new pole}$ within the strip $-\pi \leq \Im m \, \theta \leq 0$ in the complex $\theta$-plane. Obviously,
it has to be accompanied by a new zero at $\theta=\theta_\text{new zero}$, close to the new pole, $\theta_\text{new zero}-
\theta_\text{new pole}\sim \eta$, so that the pole disappears at $\eta=0$. The position of the new pole can't be deep inside
the above strip, because then the associated zero would lead to a pole within the physical strip $0 < \Im m\, \theta < \pi$, as
dictated by the first equation in \eqref{S11crossunitarity}\footnote{For more details on possible locations of poles and
zeroes of the elastic 2 $\to$ 2 amplitude see e.g. \cite{Zamolodchikov:2011wd,zamolodchikov2013ising,Gabai:2019ryw}}. This would imply appearance of a singularity in the upper half-plane of the complex energy, incompatible with the physical requirement of macro causality \cite{iagolnitzer1981analyticity} (the statement that the scattering amplitudes can not have singularities on the principal sheet of the energy surface). Therefore, the new pole can only appear close to the boundary $\Im m \,\theta = 0$ or $\pi$, so that the associated new zero
would be located in the physical strip $0 < \Im m \,\theta < \pi$. In this case the pole $\theta_\text{new pole}$ would
signal a conventional resonance state, with the width proportional to the separation between the new pole and
and the associated new zero. This interpretation immediately rules out possibility of emergence of the new poles in the order $\eta$,
since then by continuation to the opposite sign of $\eta$ one could make the width negative, in contradiction with unitarity \footnote{Resonance poles can appear in $S_{11}(\theta)$, and indeed do appear, in the order $\eta^2$, see e.g.\cite{Delfino:2005bh}.}.
To summarize, no additional poles in the finite part of the $\theta$-plane can appear in $\Phi^{(1)}(\theta)$.

Another possible objection concerns asymptotic $\theta \to \infty$ behavior of $S_{11}(\theta)$ in the order $\sim\eta$.
Without violation of crossing and unitarity one can add to \eqref{S11form} an entire function of $\theta$, subject to the
conditions \eqref{S11crossunitarity}. Since those conditions suggest $2\pi i$ periodicity of the amplitude, such additional entire function could be either a constant (impossible, as it would violate the condition $S_{11}(0)=-1$), or grow at least
as $\sinh\theta$ as $\theta\to \infty$. However, analysis of the form factor perturbation theory suggested that
the corrections $\Phi^{(k)}(\theta)$ decay as $\theta^{k-1}\,e^{-\theta}$ as $\theta\to+\infty$; in particular
$\Phi^{(1)}(\theta)$ behaves as $e^{-\theta}$ in this limit, as \eqref{varphi1} obviously does. Indeed,
as $S_{11}(\infty)=+1$, at high energies the particles $A_1$ behave much like free bosons. In particular, the form factors
$\langle A_1(\theta_1)...A_1(\theta_N)\mid O_R(0) \mid A_1(\theta_1')...A_1(\theta_M')\rangle$ of any relevant field operator
$O_R(x)$ are bounded by a constant; the above asymptotic behavior of $\Phi^{(n)}(\theta)$ follows.

\section{Finite-Size Corrections to Mass Gap}
Consider a generic 2d massive QFT, and denote $A_a$ the stable particles of the theory, and $M_a$ their masses. We assume that $A_1$ is the lightest particle in the spectrum. If one puts this theory in the geometry of a cylinder $x \sim x+R$, the energy spectrum acquires the finite size corrections. Thus, the leading large-$R$ correction to the ground state energy is
\begin{eqnarray}
E_0(R) = F R - \sum_a M_a \int \frac{d\theta}{2 \pi} e^{-M_a R \cosh\theta} \cosh \theta + \cdots  \,,
\end{eqnarray}
where $F$ is the bulk vacuum energy density. Furthermore, if $E_1(R)$ is the energy of the state of one particle $A_1$ at rest (See e.g. \cite{Klassen:1990ub})
\begin{eqnarray}
E_1(R) = F R + M_1 + \sum_{b,c} M_1 \kappa^1_{bc} e^{- \mu^1_{bc} R}
- \sum_a M_a \int \frac{d\theta}{2 \pi} e^{-M_a R \cosh\theta} S_{1a}(\theta + \frac{\pi i}{2})  \cosh \theta  \,, \quad \label{MassGapExpansion2}
\end{eqnarray}
The sum in the first line here includes particles $a,b$ whose masses satisfy the inequality $M^2_1 \ge |M^2_a-M^2_b|$, and
\begin{eqnarray}
\mu^a_{bc} = \frac{M_b M_c}{M_a}\sin u^a_{bc}\,, \qquad \kappa^a_{bc} = -(\Gamma^a_{bc})^2\,\mu^a_{bc} / M_a \,,
\end{eqnarray}
Here $i u_{ab}^c$ is the rapidity position of the pole associated with $A_c$ in the amplitude $S_{ab}(\theta)$ of the elastic process $A_a+A_b \to A_a + A_b$,
\begin{eqnarray}
M^2_c = M^2_a + M^2_b + 2 M_a M_b\cos u^c_{ab} \,,
\end{eqnarray}
and
\begin{eqnarray}
\big( \Gamma^c_{ab} \big)^2 = -i \underset{\theta \to i u^c_{ab}}{\text{Res}} S_{ab}(\theta)\,.
\end{eqnarray}
In the $E_8$ theory in Sec.5 we have $M_2 = 2 M_1\cos\frac{\pi}{5}$ and $M_3 = 2 M_1\cos\frac{\pi}{30}$, and
\begin{eqnarray}
\mu^1_{22} = \frac{\sqrt{5 + 2\sqrt{5}}}{2} M_1 \approx 1.53884 \, M_1 \,, \quad  \mu^1_{33} = \frac{1}{2}\sqrt{7 + 8\cos\frac{\pi}{15}} M_1 \approx 1.92517 \, M_1 \,. \quad
\end{eqnarray}
This leads to the fitting formula \eqref{MassGapExpansion3terms}.

%\section{Errata of \cite{Xu:2022mmw}: on $M_1$ singular expansion}

\bibliographystyle{unsrt}

\bibliography{isingvertexref}

\end{document}